\documentclass[journal]{IEEEtran}[12pt]
\ifCLASSINFOpdf
\usepackage[pdftex]{graphicx}
\graphicspath{{../pdf/}{../jpeg/}}
\DeclareGraphicsExtensions{.pdf,.jpeg,.png}
\else
\usepackage[dvips]{graphicx}
\graphicspath{{../eps/}}
\DeclareGraphicsExtensions{.eps}
\fi\usepackage{graphicx}

\usepackage{graphics}
\usepackage{epsfig}
\usepackage{epstopdf}
\usepackage{stfloats}
\usepackage[cmex10]{amsmath}
\usepackage{algorithmic}
\usepackage[ruled]{algorithm2e}
\usepackage{indentfirst}
\usepackage{array}
\usepackage{mdwmath}
\usepackage{mdwtab}
\usepackage{graphicx}
\usepackage{subfigure}
\usepackage{color}
\usepackage{amsfonts,amssymb}
\usepackage{multirow}
\usepackage{diagbox}
\usepackage{caption}
\usepackage{verbatim}
\usepackage{float}
\usepackage{graphicx}
\usepackage{bm}
\usepackage{bbding}
\usepackage{tabu} 
\usepackage{multirow} 
\usepackage{multicol} 
\usepackage{multirow} 
\usepackage{float} 
\usepackage{makecell}
\usepackage{booktabs} 
\usepackage [mathscr] {euscript}

\usepackage{hyperref} %
\usepackage{lineno} 

\usepackage{amsfonts,amsthm,array} 

\usepackage{amsfonts,amsthm,array} 
\makeatletter
\renewcommand{\maketag@@@}[1]{\hbox{\m@th\normalsize\normalfont#1}}
\makeatother

\newcommand{\figwidThree}{0.32 \textwidth}
\newcommand{\figheiThree}{0.24 \textwidth}

\SetKwRepeat{Do}{do}{while}

\begin{document}
	

\title{Resource Allocation for Secure Dual-UAV-Assisted ISAC System \thanks{Manuscript received.}}

\author{Hongjiang~Lei, 
	Jianshuo~Geng, 
	Ki-Hong~Park, \\
	Jia~Ye, 
	Liang~Yang, 
	Xiaqing~Miao, 
	and 
	Gaofeng~Pan
	\thanks{
		(Corresponding author: \textit{Hongjiang Lei}.)}
	\thanks{Hongjiang~Lei and Jianshuo~Geng are with School of Communications and Information Engineering, Chongqing University of Posts and Telecommunications, Chongqing 400065, China  (e-mail: leihj@cqupt.edu.cn, cquptwyy@163.com).}
	\thanks{Ki-Hong~Park is with the CEMSE Division, King Abdullah University of Science and Technology (KAUST), Thuwal 23955-6900, Saudi Arabia (e-mail: kihong.park@kaust.edu.sa).}
	\thanks{Jia~Ye is with State Key Laboratory of Power Transmission Equipment Technology, School of Electrical Engineering, Chongqing University, Chongqing 400044, China (e-mail: jia.ye@cqu.edu.cn).}
	\thanks{Liang~Yang is with the College of Computer Science and Electronic Engineering, Hunan University, Changsha 410082, China, and also with School of Information Engineering, Changsha Medical University, Changsha 410219, China. (e-mail: liangy@hnu.edu.cn).}
	\thanks{Xiaqing~Miao and Gaofeng~Pan are with the School of Cyberspace Science and Technology, Beijing Institute of Technology, Beijing 100081, China (e-mail: gfpan@bit.edu.cn).}
}

\maketitle
\begin{abstract}
	Integrated Sensing and Communication (ISAC) is an emerging technology in next-generation wireless communication  networks, enabling the simultaneous execution of communication and sensing tasks by fully utilizing limited spectrum resources. In this work, we investigate the secrecy performance of a dual-Uncrewed Aerial Vehicle (UAV)-assisted secure ISAC system. Specifically, a base station UAV communicates with users and transmits radar signals to locate potential eavesdroppers, while simultaneously providing information to a jammer UAV to perform jamming tasks. Considering constraints such as maximum UAV velocity, transmit power, propulsion energy, and sensing thresholds, we maximize the average secrecy rate by optimizing user scheduling strategies, time allocation, transmit power, and UAV trajectories. The resulting non-convex problem, originating from tightly coupled variables, is tackled by an efficient iterative algorithm. In particular, the original optimization problem is decomposed into six subproblems, and non-convex subproblems are transformed into approximately convex forms via successive convex approximation. Then, block coordinate descent techniques are employed to solve all subproblems sequentially. Numerical results demonstrate the convergence and effectiveness of the proposed algorithm.
\end{abstract}

\begin{IEEEkeywords}
	Uncrewed aerial vehicles, 
	integrated sensing and communication, 
	beamforming, 
	physical-layer security
\end{IEEEkeywords}

\section{Introduction}
\label{sec:Introduction}

\subsection{Background and related works}

\begin{table*}[t]
	\centering
	\begin{minipage}{1\textwidth}
		\captionsetup{justification=raggedright,singlelinecheck=false}
		\caption{Related Works on UAV-ISAC Systems}
		\label{table1}
		\resizebox{\linewidth}{!}{
			\begin{tabular}{c c c c c c c}
				\toprule
				\textbf{Reference}      &\textbf{Multi-UAV}       &\textbf{TD-ISAC}   &\textbf{PLS}    &\textbf{Imperfect CSI}  &\textbf{Optimization objectives} &\textbf{Main parameters}\\    
				\midrule
				\cite{LiuZ2025TVT}&   & \checkmark &    &   & Sum sensing rate &\makecell{Beamformer, transmit power, and trajectory }\\
				\cite{GangY2025DCN}& &   & & & ACR & \makecell{Trajectory, user scheduling, and beamformer} \\
				\cite{KhaliliA2024TWC}& & \checkmark & &   & \makecell{Average power \\consumption} & \makecell{Beamformer, sensing power, and time allocation}\\
				\cite{ZhouL2026TWC}&\checkmark & &  & & CSINR &\makecell{Trajectory, user scheduling, and transmit power}\\
				\cite{ChengG2025TCOM}& \checkmark & & &  & ASR &\makecell{Trajectory, user scheduling, and beamforming}\\
				\cite{XuY2024TWC}& & & \checkmark & & Sum SR &\makecell{Beamformer, AN covariance matrix, and IRS phase}\\
				\cite{ZhangJ2024TWC}& &  & \checkmark & &   ACR &\makecell{User scheduling,transmit power, and IRS phase}\\
				\cite{WangC2026TCE}&\checkmark& &\checkmark & & ACR &\makecell{Beamformer, transmit power, and IRS phase}\\
				\cite{WangY2026IoT}&  &  & \checkmark & \checkmark & ASR &\makecell{Trajectory, user scheduling, and beamformer}\\
				\cite{YaoJ2025WCL}& & & \checkmark & & ASR &\makecell{Beamformer and sensing covariance matrix}\\
				\cite{JinH2026IoT}& & &  & \checkmark & Communication rate &\makecell{Trajectory and beamformer}\\
				\cite{JinH2026TCOM}& & &  & \checkmark & SR &\makecell{Trajectory and beamformer}\\
				\cite{JiangC2025IoT}& &  & \checkmark  & \checkmark &CSINR &\makecell{Trajectory, beamformer, and user scheduling}\\
				\cite{LouY2026TVT}& \checkmark &  & \checkmark   & & ASR & \makecell{Trajectory, beamformer, and user scheduling}\\
				\cite{LiuY2024TVT}& \checkmark & \checkmark & \checkmark  & & ASR & \makecell{Trajectory, beamformer, and user scheduling}\\
				Our Work&\checkmark&\checkmark&\checkmark 	& \checkmark 	& ASR & \makecell{Trajectory, beamformer, \\user scheduling, time allocation, and transmit power}\\
				\bottomrule
			\end{tabular}
		}
	\end{minipage}
\end{table*}

Integrated Sensing and Communication (ISAC) technology breaks down the traditional boundaries between sensing and communication by sharing hardware, spectrum, and signal processing resources, which not only enables communication systems to identify and analyze channel characteristics while proactively transmitting information and perceiving surrounding physical features, thereby realizing mutual enhancement between communication and sensing, but also significantly improves spectrum and energy efficiency \cite{ZhangD2026Surveys}. Uncrewed Aerial Vehicles (UAVs), leveraging their high mobility, aerial perspective, ability to overcome terrain limitations, flexible deployment, and precise sensing capabilities, are poised to become critical nodes in next-generation wireless networks \cite{JinH2026npj}. The integration of UAVs with ISAC technology leverages their aerial advantages to overcome the fixed-coverage limitations of ground-based sensing and communication, enabling more flexible and broader range of services \cite{FeiZ2023Magzine}. Meanwhile, the strong Line-of-Sight (LoS) links between aerial platforms and the ground help improve sensing resolution and communication rates, and sensing results can further assist in beam tracking, trajectory planning, and resource allocation, thereby enhancing communication performance \cite{MengK2024WC}. 

Existing studies have extensively explored various aspects of UAV-ISAC systems. Ref. \cite{LiuZ2025TVT} proposed a UAV-assisted Time-Division ISAC (TD-ISAC) framework where the UAV periodically senses ground targets and transmits the acquired information to terrestrial Communication Users (CUs) using space division multiple access, introducing a Cramér–Rao Bound (CRB)-based radar estimation rate to quantify sensing performance and maximizing the sum sensing rate via joint optimization of time slot allocation, transmit beamforming, and UAV trajectory. Ref. \cite{JiangY2025TWC} analyzed the network-level performance of an orthogonal frequency-division multiplexing system, where ISAC-enabled Ground Base Stations (GBSs) following a two-dimensional homogeneous Poisson point process serve terrestrial CUs while sensing aerial targets, deriving analytical expressions for area communication coverage probability, area communication spectral efficiency, area radar detection coverage probability, and average CRB, and examining the impact of GBS density and height on both functionalities. Ref. \cite{GangY2025DCN} investigated a UAV-assisted full-duplex ISAC system, where the UAV integrates sensing and communication by receiving uplink communication signals and target echoes while concurrently communicating with downlink users and detecting targets, aiming to maximize the Average Communication Rate (ACR) for both uplink and downlink users by jointly optimizing user scheduling, transceiver beamforming, and UAV trajectory. 
Complementing these efforts, Ref. \cite{KhaliliA2024TWC} addressed the joint resource allocation and trajectory design problem in a multi-user, multi-target UAV-enabled TD-ISAC system with a limited backhaul link capacity between the GBS and the UAV. To avoid interference, sensing and communication were scheduled in orthogonal time slots, and the UAV trajectory, velocity, communication beamforming, sensing power, and hovering time were jointly optimized to minimize UAV power consumption while guaranteeing both communication Quality of Service (QoS) and successful sensing. 

Existing literature has also explored multi-UAV-aided ISAC systems from various perspectives. Specifically, Ref. \cite{ZhouL2026TWC} studied a multi-UAV ISAC system in which multiple UAVs simultaneously performed communication tasks for CUs and sensing tasks for Sensed Targets (STs) distributed across a specified area. To strike a balance between communication performance and sensing accuracy, a weighted optimization problem was formulated to maximize the Communication Signal-to-Interference-plus-Noise Ratio (CSINR) while minimizing the squared position error bound of STs by jointly optimizing user association, channel assignment, power allocation, and UAV deployment. In a related yet distinct direction, Ref. \cite{ChengG2025TCOM} investigated a networked ISAC architecture in which multiple GBSs cooperatively transmitted ISAC signals to communicate with several authorized UAVs while concurrently detecting unauthorized objects in a Three-Dimensional (3D) region of interest. The coordinated transmit beamforming of the GBSs, the trajectory control of the authorized UAVs, and their association with the GBSs were jointly designed to enhance the communication performance of the authorized UAVs under sensing-quality constraints.

Although the LoS links between UAVs and the ground nodes enhance communication and sensing performance, they also make signals more susceptible to eavesdropping, thereby increasing the risk of information leakage. Furthermore, the mutual interference between communication and sensing functions poses additional security challenges \cite{YangH2025survey}. To address these challenges, various secure transmission schemes have been proposed. 
Specifically, Refs. \cite{XuY2024TWC, ZhangJ2024TWC, WangC2026TCE} investigated secure Intelligent Reflecting Surface (IRS)-assisted UAV-ISAC systems. 
Ref. \cite{XuY2024TWC} introduced Artificial Noise (AN) to suppress an aerial eavesdropper while jointly optimizing UAV deployment, IRS passive beamforming, and AN power to maximize the sum Secrecy Rate (SR). Ref. \cite{ZhangJ2024TWC} considered an eavesdropper with unknown Channel State Information (CSI) and maximized the ACR by jointly designing power allocation, user/target scheduling, IRS phase shifts, and UAV trajectory. 
Ref. \cite{WangC2026TCE} proposed a joint beamforming strategy to maximize the ACR of legitimate UAVs under sensing and power constraints. Unlike existing Physical-Layer Security (PLS) studies that focus on secure communication performance, the security work in \cite{WangC2026TCE} emphasizes incorporating airspace intrusion sensing as a critical constraint in communication system design, thereby serving as a valuable complement to current PLS-oriented research. 
Moving from IRS-assisted architectures to communication-centric secure designs, Ref. \cite{WangY2026IoT} maximized the Average SR (ASR) in a UAV-assisted ISAC system by jointly optimizing user scheduling, transmit beamforming, receive filtering, and UAV trajectory. Ref. \cite{YaoJ2025WCL} addressed a scenario in which an eavesdropper simultaneously wiretapped both communication and sensing signals, maximizing the ASR via joint trajectory and beamforming design, subject to sensing and security constraints. To further enhance situational awareness, Ref. \cite{JinH2026IoT} incorporated an Extended Kalman Filter (EKF) into a UAV-assisted secure ISAC system to track and jam eavesdroppers, maximizing the uplink communication rate through joint beamforming and trajectory optimization under practical energy and velocity constraints. 
Building upon this, Ref. \cite{JinH2026TCOM} developed an EKF-assisted joint trajectory prediction and tracking scheme for both legitimate and eavesdropping users, and jointly optimized transmit beamforming and UAV trajectory to maximize the SR.
Ref. \cite{JiangC2025IoT} studied a secure integrated sensing, communication, and computing system, where a UAV suppressed a potential eavesdropper while providing offloading services to ground CUs, minimizing total energy consumption of CUs via joint optimization of offloading ratios, user scheduling, beamforming, and UAV trajectory. Ref. \cite{LouY2026TVT} considered a dual-UAV-assisted system in which an ISAC UAV and a jamming UAV operated cooperatively in a hybrid monostatic-bistatic radar configuration. The jamming UAV emitted AN to help the source UAV detect multiple ground targets while simultaneously interfering with an eavesdropper. Considering imperfect Successive Interference Cancellation (SIC) and constraints on the transmit power budget, UAV maneuverability, and sensing requirements, the ASR was maximized by jointly optimizing the dual-UAV trajectories and beamforming. 
Ref. \cite{LiuY2024TVT} investigated the secrecy performance of an aerial TD-ISAC system with multiple ground users and targets, and the sum SR was maximized by jointly designing user scheduling, transmit power, and UAV trajectory. 
{The key distinctions of this work from previous UAV-ISAC systems are summarized in Table \ref{table1}.}

\subsection{Motivation and contributions}

Despite considerable progress in UAV-assisted ISAC, most existing works rely on idealized assumptions, e.g., perfect CSI, while ignoring practical constraints such as limited propulsion energy, realistic sensing thresholds, and imperfect eavesdropper location knowledge. This motivates us to propose a practical and holistic secure ISAC framework, where a Base-Station (BS) UAV and a friendly jammer UAV cooperatively execute communication, sensing, and jamming tasks under realistic conditions. In this context, we investigate two key problems in a dual-UAV-assisted TD-ISAC system: \textit{How to jointly design trajectories and beamforming to optimize secrecy performance, and how to enable reliable transmission of the sensed information about the eavesdropper from the BS UAV to the jammer UAV, thereby ensuring effective jamming during confidential data delivery}. To resolve these problems, we thoroughly analyze the performance of the considered system with multiple legitimate users and one eavesdropper, and develop a closed-loop framework that unifies sensing, jamming, and communication. The main contributions are summarized as follows.
\begin{enumerate}
	\item We consider a secure dual-UAV-assisted ISAC system, where one UAV acts as an aerial BS that serves ground users via downlink communications while simultaneously transmitting radar signals to detect an eavesdropper. The detected information is forwarded to the friendly aerial jammer, which then broadcasts AN to suppress the eavesdropper. A user scheduling scheme is adopted to mitigate mutual interference while ensuring communication fairness. Under constraints on initial/final positions, maximum speed, transmit power, energy consumption, and sensing thresholds, we maximize the ASR by jointly optimizing user scheduling, time allocation, transmit power, sensing beamforming, and UAV trajectories.
	
	\item Unlike Refs.~\cite{LiuZ2025TVT, KhaliliA2024TWC}, which investigated TD-ISAC systems without security considerations, our work addresses the challenging scenario of encountering an eavesdropper with location uncertainty. Specifically, we optimize user scheduling, time allocation, transmit power, UAV trajectories, and beamforming to enhance secrecy performance in a dual-UAV-assisted TD-ISAC system.
	
	\item Compared with Ref.~\cite{LiuY2024TVT}, which studied secrecy performance in TD-ISAC under perfect eavesdropper location information, our work considers a more practical setting with imperfect location knowledge and a multi-antenna jamming UAV. In addition, we jointly optimize time allocation and beamforming, and further analyze the impact of location errors on secrecy performance.
\end{enumerate}

The rest of this paper is organized as follows. The system model and problem formulation are provided in Section \ref{sec:SystemModel}. Section \ref{sec:proposed solution} presents the joint algorithm for user scheduling, time alllocation, transmit beamforming, and UAV trajectory optimization. Simulation results are shown in Section \ref{sec:Simulation Results}. Finally, Section \ref{sec:Conclusion} concludes this paper. 
{A summary of the symbols and their definitions used in this work is provided in Table~\ref{table2}.}

\begin{table}[t]
	\centering
	\begin{minipage}{0.4\textwidth}
		\captionsetup{justification=raggedright,singlelinecheck=false}
		\caption{Symbols and Description}
		\label{table2}
		\resizebox{\linewidth}{!}{
			\begin{tabular}{c l}
				\toprule
				{Notation}  & {Description}				\\
				\midrule
				$U$, $J$, $U_k$, $E$ & BS UAV, Jamming UAV, the $k$-th user, and eavesdropper \\
				$K$, $N_t$ & Number of users and antennas on $J$\\
				$T$, $N$, $\delta$ & Total flight period, number of slots, and duration of a slot \\
				$\mathbf{q}_i\left[ n \right]$ & Position of UAV $i$\\
				$\mathbf{q}_k$, $\mathbf{q}_E$ & Position of $U_k$ and $E$\\
				${{\mathbf{v}}_{i}}\left[ n \right]$ & Horizontal velocity of UAV $i$ \\
				$\mathbf{w}_J\left[ n \right]$ & Beamforming at UAV $J$ \\
				${\alpha _k}\left[ n \right]$ & User scheduling variable \\
				${P_s}\left[ n \right]$ & The effective sensing power \\
				$P_{i,{\mathrm{ave}}}^{{\mathrm{fly}}}$ & Average propulsion power of UAV $i$ \\
				${P_{i, \max }}$ & The maximum transmit power of UAV $i$ \\
				${V_{\max }}$ and ${a_{\max }}$ & The maximum velocity and acceleration of the UAVs\\
				${R_{\min }^{{\mathrm{sen}}}}$ & The minimum sensing threshold \\
				${R _{\min }^{{\mathrm{sec}}}}$ & The minimum communication requirement \\
				${\eta}\left[ n \right]$ & The time allocation ratio \\
				\bottomrule
			\end{tabular}
		}
	\end{minipage}
\end{table}

\section{System model and problem formulation}
\label{sec:SystemModel}

\subsection{System model}

\begin{figure}[t]
	\centering
	\includegraphics[width = 0.4 \textwidth]{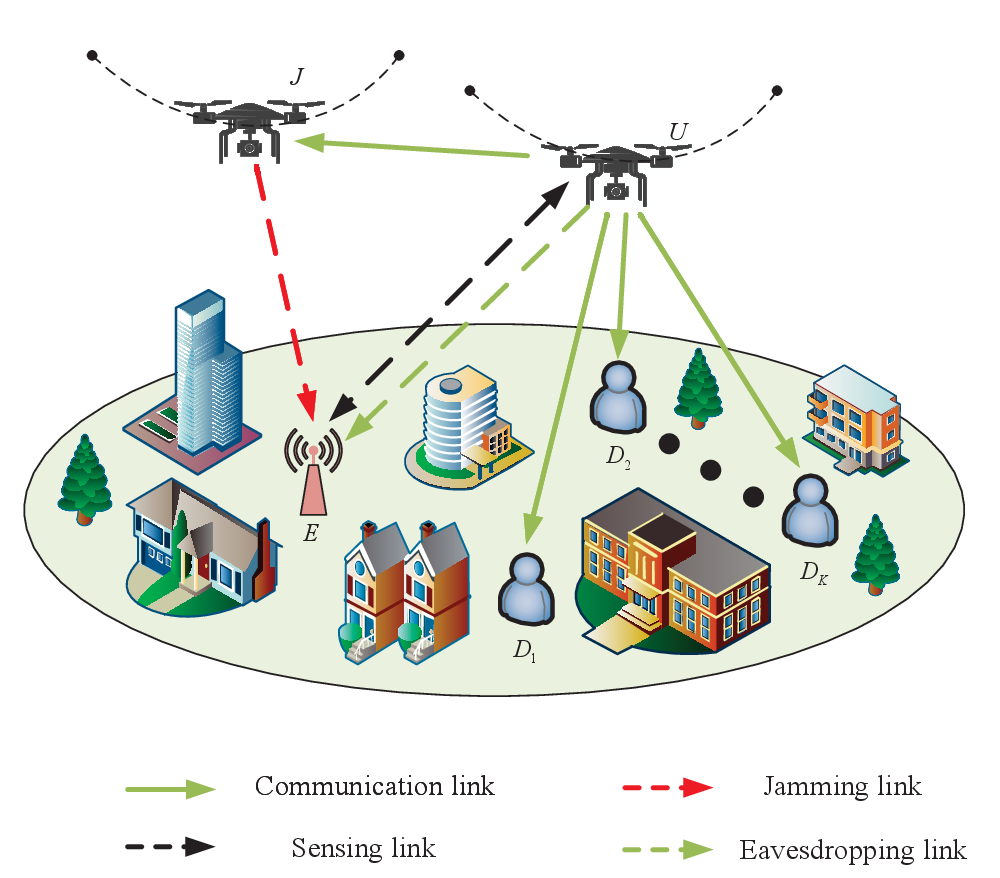}
	\caption{System model.}
	\label{moxing}
\end{figure}

\begin{figure}[t]
	\centering
	\includegraphics[width = 0.35 \textwidth]{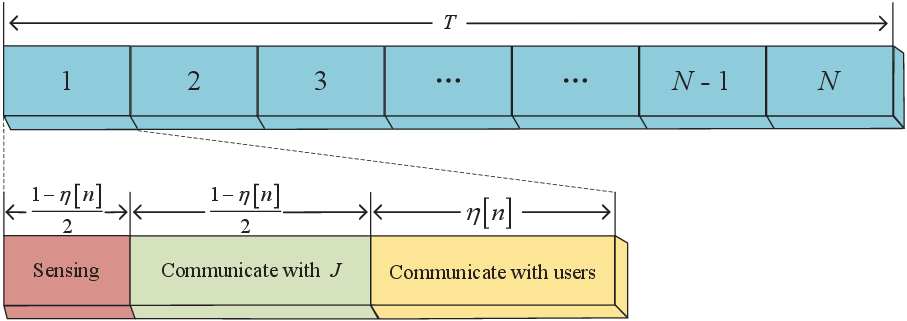}
	\caption{TD-ISAC frame structure.}
	\label{TimeSlot}
\end{figure}

As shown in Fig. \ref{moxing}, a dual-UAV-aided ISAC system is considered. The BS UAV $\left(U\right)$ communicates with $K$ ground users $\left( {{D_k},k = 1, \ldots, K} \right)$, while simultaneously transmitting sensing signals to estimate the location of the ground eavesdropper $\left(E\right)$. The estimated location information is transmitted to a friendly aerial jammer $\left( J \right)$ that transmits AN using a Uniform Linear Array (ULA) to suppress $E$. 
For ease of analysis, the flight period $T$ is discretized into $N$ time slots, each of duration ${\delta} = T/N$. When $N$ is sufficiently large, the position of each UAV can be assumed to be constant within each time slot 
\cite{LeiH2024TCCN, LeiH2026TCCN, MHua2024TWC, LuW2022TCOM, DingY2023JSTSP}.   
In the $n$-th time slot, the horizontal position of UAV $i\left( {i \in \left\{ {U,J} \right\}} \right)$ is denoted as $\mathbf{q}_i\left[ n \right] = \left[ x_i\left[ n \right], y_i\left[ n \right] \right]^T$, and its altitude is expressed as $H_i$. 
Similar to \cite{MHua2024TWC, WangY2026IoT, DanQ2025TVT}, the location of $E$ is assumed to be obtained via preliminary coarse estimation and is therefore prone to estimation errors. 
Specifically, $\mathbf{q}_E$ lies within an uncertainty region defined as $\mathbf{q}_E^0 = \left\{ {{{\left[ {{x_E},{y_E}} \right]}^T} \mid {x_E} \in \left[ {{{\tilde x}_E} - \Delta ,{{\tilde x}_E} + \Delta } \right], {y_E} \in \left[ {{{\tilde y}_E} - \Delta ,{{\tilde y}_E} + \Delta } \right]} \right\}$, where $\left( {{{\tilde x}_E},{{\tilde y}_E}} \right)$ represents the center coordinates of the region, and $\Delta$ denotes the half-width of the region\footnote{
	
	In this work, the location of $E$ is modeled in a simplified manner. Future work will focus on developing a data-driven, robust optimization framework, as in \cite{DingX2026TVT}. The exact eavesdropper location and its error range will be used as conditional inputs, and an adversarial training mechanism will be leveraged to learn and generate the distribution of the ``worst-case'' eavesdropper locations within this uncertain region. This data-driven approach can more accurately capture the extreme impact of location uncertainty, thereby addressing a robust optimization problem that maximizes the worst-case SR.
}. 
\color{black}
The coordinates of $D_k$ are denoted as $\mathbf{q}_k = \left[ x_k, y_k \right]$. 
Similar to \cite{LiuY2024TVT,NingZ2026JSAC}, each time slot is divided into sub-slots, shown in Fig. \ref{TimeSlot}. In the first sub-slot, $U$ senses the eavesdropper to obtain the results, which are transmitted to the jammer UAV $J$ in the second sub-slot. In the third sub-slot, $U$ communicates with the users and $J$ transmits the AN  to suppress $E$. The time allocation ratios for sensing, communication with $J$, and communication with users are $\frac{{1 - \eta \left[ n \right]}}{2}$, $\frac{{1 - \eta \left[ n \right]}}{2}$, and ${\eta}\left[ n \right]$, respectively, {where ${\eta}\left[ n \right]$ denotes the time splitting ratio for the communication sub-slot.}

\subsection{Communication model}

{
	Similar to \cite{LiuZ2024TWC}, the Air-to-Ground (A2G) channels are modelled as a LoS link without small-scale fading  \cite{JiangL2026IoT}\footnote{
		{
			It is worth noting that small-scale fading is not considered in this work. However, by following the approach in \cite{DanQ2025TCCN}, which approximates random variables using their expectations, our results can be readily extended to scenarios involving small-scale fading.  		
		}
	}. 
	The signal transmitted by $U$ is $s\left[ n \right]$ and $\mathbb{E}\{ |s\left[ n \right]|^2 \} = 1$. The channel gains from  $U$ to terrestrial receivers $r$ $\left( {r \in \left\{ {k, E} \right\}} \right)$ are expressed as\footnote{
		It is worth emphasizing that the single-antenna assumption on the source UAV is adopted in accordance with the TD-ISAC architecture, where sensing and communication are separated in different sub-slots, as shown in Fig. \ref{TimeSlot}. In this case, beamforming is not required at the source UAV, and a single antenna is sufficient for both sub-slots. The extension to a multi-antenna source UAV with joint beamforming is an important direction and is studied in our complementary work.
}}
\begin{align}
	{h_{Ur}}\left[ n \right] = \frac{{{\rho _{\mathrm{com}}}}}{{d_{Ur}^2\left[ n \right]}}
	\label{hUr}
\end{align}
where 
${{\rho_{\mathrm{com}}}}$ is the channel gain at a reference distance of 1 m
and
${d_{Ur}}\left[ n \right] = \sqrt {{H_U}^2 + {{\left\| {{{\mathbf{q}}_U}\left[ n \right] - {{\mathbf{q}}_r}} \right\|}^2}}$, which denotes the distance between $U$ and $r$.
The achievable rate at $D_k$ is expressed as 
\begin{align}
	{R_{Uk}}\left[ n \right] = {\eta}\left[ n \right]{\log _2}\left(1 + {\gamma _k}\left[ n \right]\right)
	\label{}
\end{align}
where 
${\gamma _k}\left[ n \right] = \frac{{{P_U}\left[ n \right]{h_{Uk}}\left[ n \right]}}{{{\sigma ^2}}}$, 
${P_U}\left[ n \right]$ is the transmit power of $U$, 
and 
${\sigma ^2}$ is the power of Additive White Gaussian Noise (AWGN). 

Similar to \cite{ChengG2025TCOM}, the channel between $J$ and $E$ is expressed as
\begin{align}
	{{\mathbf{h}}_{JE}}\left[ n \right] = {{\mathbf{a}}_E}\left[ n \right]\sqrt {\frac{{{\rho _{\mathrm{com}}}}}{{{d_{JE}}{{\left[ n \right]}^2}}}}
	\label{}
\end{align}
where 
${{\mathbf{a}}_E}\left[ n \right] = {\left[ {1,{e^{j\frac{{2\pi }}{\lambda }d\cos {\theta _E}\left[ n \right]}}, \ldots ,{e^{j\frac{{2\pi }}{\lambda }d({N_t} - 1)\cos {\theta _E}\left[ n \right]}}} \right]^T}$ is the steering vector, 
$N_t$ denotes the number of antennas on $J$, 
${\theta _E}$ denotes the angle of departure from $J$ to $E$, 
and 
${d_{JE}}\left[ n \right] = \sqrt {{H_J}^2 + {{\left\| {{{\mathbf{q}}_J}\left[ n \right] - {{\mathbf{q}}_E}} \right\|}^2}}$ denotes the distance between $J$ and $E$.
The signal transmitted by $J$ is expressed as
\begin{align}
	{x_J}\left[ n \right] = {{\mathbf{w}}_J}\left[ n \right]{s_{{\mathrm{AN}}}}\left[ n \right]
	\label{}
\end{align}
where 
${{\mathbf{w}}_J}\left[ n \right] \in {\mathbb{C}^{{N_t} \times 1}}$ represents the transmit beamforming vector\footnote{It is assumed that the jamming signals transmitted by $J$ can be a Gaussian pseudo-random sequence or
	utilize the deterministic waveforms similar to the structure of
	the desired signal as \cite{LvL2019, CaiY2018, XingH2016}. Then, the jamming
	signals sent by $J$ can be canceled by the received signal at all legitimate receivers.
}, $s_{\mathrm{AN}}\left[ n \right]$ denotes the jamming signal and $\mathbb{E}\{ |s_{\mathrm{AN}}\left[ n \right]|^2 \} = 1$. 
The signal received by $E$ is expressed as
\begin{align}
	{x_E}\left[ n \right] & =  {\mathbf{h}}_{JE}^H\left[ n \right]{{\mathbf{w}}_J}\left[ n \right]{s_{{\mathrm{AN}}}}\left[ n \right] + {\left( {h_{UE}^H\left[ n \right]{P_U}\left[ n \right]} \right)^{1/2}}s\left[ n \right] + {n_E}\left[ n \right]
\end{align}
where 
${n_E}\left[ n \right] \sim {\cal N}\left( {0,{\sigma ^2}} \right)$ represents AWGN.
The eavesdropping rate at $E$ is expressed as
\begin{align}
	{R_E}\left[ n \right] = {\eta}\left[ n \right]{\log _2}(1 + {\gamma _E}\left[ n \right])
	\label{}
\end{align}
where 
${\gamma _E}\left[ n \right] = \frac{{{P_U}\left[ n \right]{h_{UE}}\left[ n \right]}}{{{\sigma ^2} + {\mathrm{tr}}\left( {{{\mathbf{H}}_{JE}}\left[ n \right]{\mathbf{W}}\left[ n \right]} \right)}}$, 
${{\mathbf H}_{JE}}\left[ n \right] = {{\mathbf{h}}_{JE}}\left[ n \right]{{\mathbf{h}}_{JE}}{\left[ n \right]^H}$, and ${\mathbf{W}}\left[ n \right] = {{\mathbf{w}}_J}\left[ n \right]{{\mathbf{w}}_J}{\left[ n \right]^H}$. 

Similar to Ref. \cite{LiuY2024TVT}, it is assumed that $J$ utilizes a single receive antenna to receive the sensing message. The air-to-air channel is assumed to be LoS and the channel gain from $U$ to $J$ is expressed as
\begin{align}
	{h_{UJ}}\left[ n \right] = \frac{{{\rho _{\mathrm{com}}}}}{{d_{UJ}^2\left[ n \right]}}
	\label{}
\end{align}
where 
${d_{UJ}}\left[ {n} \right] = \sqrt {{{\left\| {{{\mathbf{q}}_U}\left[ n \right] - {{\mathbf{q}}_J}\left[ n \right]} \right\|}^2} + {{\left( {{H_U} - {H_J}} \right)}^2}} $ is the distance between $U$ and $J$. The achievable rate at $J$ is expressed as
\begin{align}
	{R_{UJ}}\left[ n \right] = \frac{{\left( {1 - \eta \left[ n \right]} \right)}}{2}{\log _2}\left( {1 + {\gamma _J}\left[ n \right]} \right)
	\label{}
\end{align}
where 
${\gamma _J}\left[ n \right] = \frac{{{P_U}\left[ n \right]{h_{UJ}}\left[ n \right]}}{{{\sigma ^2}}}$. 

To ensure fairness among users, a binary variable ${\alpha _k}\left[ n \right]$ is introduced to represent user scheduling. When ${\alpha_k}\left[ n \right]=1$, $U$ communicates with $D_k$; otherwise, ${\alpha _k}\left[ n \right] = 0$. Assuming that $U$ serves at most one user per time slot, the constraints are formulated as
\begin{subequations}
	\begin{align}
		&\sum\limits_{k = 1}^K {{\alpha _k}} \left[ n \right] \le 1, \forall n \label{diaodu1}\\
		&{\alpha _k}\left[ n \right] \in \left\{ {0,1} \right\}, \forall k,n \label{diaodu2}
	\end{align}
\end{subequations}

To account for the location uncertainty of the eavesdropper $E$, we uniformly sample $M$ horizontal positions within the target area as candidate eavesdropper locations, denoted by $\widehat{\mathbf{q}}_{E_q}$, $q = 1, \ldots, M$. The distance between $U$ and the $q$-th sampled point is then given by $d_{U E_q}[n] = \sqrt{\| \mathbf{q}_U[n] - \widehat{\mathbf{q}}_{E_q} \|^2 + H_U^2}$. This formulation effectively transforms the original problem with an uncertain eavesdropper location into an equivalent scenario involving multiple eavesdroppers at known positions. Consequently, the achievable secrecy rate is evaluated based on the worst-case performance across the sampled set $\{\widehat{\mathbf{q}}_{E_q}\}_{q=1}^{M}$. 
Accordingly, we adopt the worst-case SR over the entire eavesdropper activity region as the final objective value. Specifically, 
the ASR of the considered system is expressed as \cite{ZhongC2019CL}\footnote{
	For clarity of presentation, the terms involving $\eta \left[ n \right]$ are factored out and treated as external multipliers for their respective equations. All subsequent formulations follow this convention.
}
\begin{align}
	{R_{\sec }} = \frac{1}{N}\sum\limits_{n = 1}^N {\sum\limits_{k = 1}^K {{\eta }\left[ n \right]{\alpha _k}\left[ n \right]{R_{\sec ,k}}\left[ n \right]} }
	\label{secRate0}
\end{align}
where 
${R_{\sec ,k}}\left[ n \right] = {\left[ {{R_{Uk}}\left[ n \right] - \mathop {\max }\limits_{{{\mathbf{q}}_E} \in \mathbf{q}_E^0} {R_{UE}}\left[ n \right]} \right]^ + }$ 
and 
${\left[ x \right]^ + } = \max \left( {0,x} \right)$.

To guarantee the users' QoS, the achievable SR for each user during the flight must exceed a minimum threshold, thereby satisfying the constraint
\begin{align}
	\sum\limits_{n = 1}^N {\eta \left[ n \right]{\alpha _k}\left[ n \right]{R_{\sec, k }}\left[ n \right]}  > {R _{\min }^{{\mathrm{sec}}}}, \forall k
	\label{QoS0}
\end{align}
where
${R _{\min }^{{\mathrm{sec}}}}$ is the minimum SR requirement.

\color{black}
\subsection{Sensing model}

{
	According to \cite{LiuX2024IOT}, the channel gain of the sensing echo link from $U$ to $E$ is expressed as
	\begin{align}
		{h_{\mathrm{sens}}}\left[ n \right] = \frac{{{\rho _{{\mathrm{sens}}}}}}{{d_{UE}^4\left[ n \right]}}
		\label{}
	\end{align}
	where ${\rho_{\mathrm{sens}}}$ is the sensing attenuation coefficient at a reference distance. 
	The sensing echo SNR of $U$ is expressed as $\gamma^{{\mathrm{sens}}}\left[ n \right] = \frac{{{P_U}\left[ n \right]{h_{{\mathrm{sens}}}}\left[ n \right]}}{{{\sigma ^2}}}$, which is utilized to evaluate the sensing performance.
	To ensure the sensing performance, we impose the following constraint
	\begin{align}
		\mathop {\min }\limits_{{{\mathbf{q}}_E} \in \mathbf{q}_E^0} {R^{{\mathrm{sens}}}}\left[ n \right] \ge {R_{\min }^{{\mathrm{sen}}}}, \forall n
		\label{}
	\end{align}
	where
	${R^{{\mathrm{sens}}}}\left[ n \right] = \frac{{\left( {1 - \eta \left[ n \right]} \right)}}{2}\log \left( {1 + \gamma _{}^{{\mathrm{sens}}}\left[ n \right]} \right)$ 
	and 
	${R_{\min }^{{\mathrm{sen}}}}$ is the minimum sensing threshold.
	Assuming that the sensing information from $J$, we impose the following constraint \cite{LiuY2024TVT}
	\begin{align}
		{R_{UJ}}\left[ n \right] \ge \mathop {\max }\limits_{{{\mathbf{q}}_E} \in \mathbf{q}_E^0} {R^{{\mathrm{sens}}}}\left[ n \right], \forall n
		\label{}
	\end{align}
	
	\subsection{Energy consumption}
	During trajectory optimization, it is essential to balance path selection and energy consumption to ensure the UAV retains sufficient battery power for subsequent operations after completing its communication tasks. 
	To maintain UAV power consumption within manageable limits, we impose the following average propulsion power constraint
	\begin{align}
		\frac{1}{N}\sum\limits_{n = 1}^N {P_i^{{\mathrm{fly}}}\left[ n \right]}  \le P_{i,{\mathrm{ave}}}^{{\mathrm{fly}}}, i \in \left\{ {U,J} \right\}
		\label{}
	\end{align}
	where 
	$P_i^{{\mathrm{fly}}}\left[ n \right] = {P_0}\left( {1 + \frac{{3\left\| {{{\bf{v}}_i}\left[ n \right]} \right\|^2}}{{U_{{\mathrm{tip}}}^2}}} \right) + \frac{1}{2}{\mkern 1mu} {d_o}{\mkern 1mu} \rho sA\left\| {{{\bf{v}}_i}\left[ n \right]} \right\|^3 + {P_1}{\left( {\sqrt {1 + \frac{\left\| {{{\bf{v}}_i}\left[ n \right]} \right\|^4}{{4v_0^4}}}  - \frac{\left\| {{{\bf{v}}_i}\left[ n \right]} \right\|^2}{{2v_0^2}}} \right)^{1/2}}$ \cite{ZengY2019TWC},  $P_{i,{\mathrm{ave}}}^{{\mathrm{fly}}}$ represents the average horizontal propulsion power of UAV $i$,  
	${{\mathbf{v}}_i}\left[ n \right] = \frac{{({{\mathbf{q}}_i}\left[ n \right] - {{\mathbf{q}}_i}[n - 1])}}{\delta }$ denotes the flight velocity of UAV $i$, 
	${P_0}$ and ${P_1}$ represent the blade profile power and rotor-induced power, respectively,  
	${U_{{\mathrm{tip}}}}$ denotes the tip speed of the rotor blade,  
	${d_0}$ and $\rho$ are the fuselage drag ratio and air density, respectively,  
	${v_0}$ is the mean rotor-induced velocity,  
	$s$ and $A$ denote to the rotor solidity and rotor disc area, respectively.
	
	\subsection{Problem formulation}

	In this paper, the ASR is maximized by jointly designing user scheduling, time allocation, transmit power, trajectories of UAVs, and the beamforming vector.  
	Define 
	${\mathbf{A}} = \left\{ {{\alpha _k}\left[ n \right], \forall k,n} \right\}$, 
	${\mathbf{Y}} =\left\{ {\eta \left[ n \right], \forall n} \right\}$,
	${\mathbf{P}} = \left\{ {{P_U}\left[ n \right], \forall n} \right\}$,  
	${\mathbf{W}} = \left\{ {{{\mathbf{w}}_J}\left[ n \right], \forall n} \right\}$,  
	${{\mathbf{Q}}_{\mathrm{U}}} = \left\{ {{{\mathbf{q}}_U}\left[ n \right], \forall n} \right\}$,  
	and 
	${{\mathbf{Q}}_{\mathrm{J}}} = \left\{ {{{\mathbf{q}}_J}\left[ n \right], \forall n} \right\}$.  
	Thus, we formulate the following problem 
	\begin{subequations}
		\begin{align}
			{{\cal P}_0}: \quad &\mathop {\max }\limits_{{\mathbf{A}},{\mathbf{P}},{\mathbf{W}}, {\mathbf{Y}},{{\mathbf{Q}}_{\mathrm{U}}},{{\mathbf{Q}}_{\mathrm{J}}}} {R_{\sec }}										\label{P0_a}\\
			\text{s.t.} \quad &\sum\limits_{k = 1}^K {{\alpha _k}\left[ n \right] \le 1, \forall n} 	\label{P0_b}\\
			&{\alpha _k}\left[ n \right] \in \left\{ {0,1} \right\}, \forall k,n 	\label{P0_c}\\
			&0 \le {P_U}\left[ n \right] \le {P_{U,\max }}, \forall n				\label{P0_d}\\
			&\frac{1}{N}\sum\limits_{n = 1}^N {P_i^{{\mathrm{fly}}}\left[ n \right]}  \le P_{i,{\mathrm{ave}}}^{{\mathrm{fly}}},i \in \left\{ {U,J} \right\}		\label{P0_E}\\
			&{\mathrm{tr}}\left( {{\mathbf{W}}\left[ n \right]} \right) \le {P_{J,\max }}, \forall n \label{P0_f}\\
			&\mathbf{W}\left[ n \right] \succeq 0,		\forall n 										\label{P0_g}\\
			&{\mathrm{rank}}\left( {{\mathbf{W}}\left[ n \right]} \right) = 1, 	\forall n	\label{P0_h}\\
			&\sum\limits_{n = 1}^N {\eta \left[ n \right]{\alpha _k}\left[ n \right]{R_{\sec, k}}\left[ n \right]}  > {R _{\min }^{{\mathrm{sec}}}}, \forall k														\label{P0_i}\\
			&{R_{UJ}}\left[ n \right] \ge \mathop {\max }\limits_{{{\mathbf{q}}_E} \in \mathbf{q}_E^0} {R^{{\mathrm{sens}}}}\left[ n \right], \forall n			\label{P0_j}\\
			&\mathop {\min }\limits_{{{\mathbf{q}}_E} \in \mathbf{q}_E^0} {R^{{\mathrm{sens}}}}\left[ n \right] \ge {R_{\min }^{{\mathrm{sen}}}},			\forall n	\label{P0_k}\\
			&{{\mathbf{q}}_{_i}}\left[ 1 \right] = {\mathbf{q}}_i^{\textrm{I}},{{\mathbf{q}}_i}\left[ N \right] = {\mathbf{q}}_i^{\textrm{F}}, \forall i \in \left\{ {U,J} \right\}									\label{P0_l}\\
			&\left\| {{{\mathbf{q}}_i}\left[ n + 1 \right] - {{\mathbf{q}}_i}\left[ n \right]} \right\| = {V_{i}}\delta , \forall n,	i \in \left\{ {U,J} \right\}													\label{P0_m}\\
			&\left\| {{\mathbf{v}_{i}}\left[ n \right]} \right\| \le {V_{\max }}, \forall n,	i \in \left\{ {U,J} \right\}			\label{P0_n}\\
			&\left\| {{\mathbf{v}_{i}}\left[ n + 1 \right] - {\mathbf{v}_{i}}\left[ n \right]} \right\| \le \delta {a_{\max }}, \forall  n,	i\in\left\{{U,J}\right\}			\label{P0_o} \\
			& 0 < {\eta}\left[ n \right] <1, \forall  n \label{P0_p}
		\end{align}
	\end{subequations}
	where 
	${P_{U, \max }}$ and ${P_{J, \max }}$ denote the maximum transmit power of $U$ and $J$, respectively, and
	${V_{\max }}$ and ${a_{\max }}$ denote the maximum velocity and acceleration constraints of the UAVs, respectively. 
	In particular, 
	Eqs. (\ref{P0_b}) and (\ref{P0_c}) represent the scheduling constraints of UAV $U$,  
	Eq. (\ref{P0_d}) denotes the transmit power of $U$,  
	Eq. (\ref{P0_E}) represents the horizontal propulsion power of the UAV,  
	Eq. (\ref{P0_f}) denotes the transmit power constraint of the beamforming vector,  
	Eq. (\ref{P0_g}) and (\ref{P0_h}) are positive semi-definite and rank-one constraints,
	Eq. (\ref{P0_i}) ensures the user QoS,
	Eq. (\ref{P0_j}) ensures that the communication rate between $U$ and $J$ is greater than the sensing rate,  
	Eq. (\ref{P0_k}) ensures the sensing quality of $U$,  
	Eq. (\ref{P0_l}) restricts the initial and final positions of the UAVs,  
	Eq. (\ref{P0_m}) limits the flight distance of the UAVs,  
	Eqs. (\ref{P0_n}) and (\ref{P0_o}) are the velocity and acceleration constraints of the UAVs, respectively,
	and 
	Eq. (\ref{P0_p}) denotes the time allocation constraint of UAV $U$.
	
	$\mathcal{P}_{0}$ is a nonlinear problem due to the following reasons. 
	First, the ${\left[  \cdot  \right]^ + }$ function is discontinuous, non-concave, and nonlinear.
	The constraints given by Eqs. (\ref{P0_b}) and (\ref{P0_c}) involve binary integer restrictions, which are non-convex. 
	Second, the variables in Eq. (\ref{P0_E}) are highly coupled, resulting in non-convexity. 
	Additionally, the constraints given by Eqs. (\ref{P0_f}), (\ref{P0_g}), and (\ref{P0_h}) are all non-convex. 
	Moreover, Eqs. (\ref{P0_j}) and (\ref{P0_k}) are non-convex due to the strong coupling of trajectories. Therefore, the optimization problem is formulated as a non-convex, nonlinear mixed-integer programming problem.

	\newcounter{TempEqCnt}
	\setcounter{TempEqCnt}{\value{equation}}
	\setcounter{equation}{23} 
	\begin{figure*}[ht]
		
		\begin{align}
			{R_{UE, 2}}\left[ n \right] &= {\log _2}\left( {{\sigma ^2} + {\mathrm{tr}}({{\mathbf{H}}_{JE}}\left[ n \right]{{\bf{W}}^{\left( l \right)}}\left[ n \right]) + {P_U}\left[ n \right]{h_{UE}}\left[ n \right]} \right) + \frac{{{\mathrm{tr}}\left( {{{\bf{H}}_{JE}}\left[ n \right]\left( {{\bf{W}}\left[ n \right] - {{\bf{W}}^{\left( l \right)}}\left[ n \right]} \right)} \right)}}{{\ln \left( 2 \right)\left( {{\sigma ^2} + {P_U}\left[ n \right]{h_{UE}}\left[ n \right] + {\mathrm{tr}}\left( {{{\bf{H}}_{JE}}\left[ n \right]{{\bf{W}}^{\left( l \right)}}\left[ n \right]} \right)} \right)}}\nonumber\\ 
			&- {\log _2}\left( {{\sigma ^2} + {\mathrm{tr}}\left( {{{\bf{H}}_{JE}}\left[ n \right]{\bf{W}}\left[ n \right]} \right)} \right)
			\label{RUEin13}
		\end{align}
		\hrulefill
	\end{figure*}
	\setcounter{equation}{26}
	\begin{figure*}[ht]
		\begin{align}
			{R_{Uk, 1}}\left[ n \right]&= {\log _2}\left( {1 + \frac{{{P_U}\left[ n \right]{\rho _{{\mathrm{com}}}}}}{{{\sigma ^2}\left( {{H_U}^2 + {{\left\| {{\bf{q}}_U^{\left( l \right)}\left[ n \right] - {{\bf{q}}_k}} \right\|}^2}} \right)}}} \right) - \frac{{{P_U}\left[ n \right]{\rho _{{\mathrm{com}}}}\left( {{{\left\| {{{\bf{q}}_U}\left[ n \right] - {{\bf{q}}_k}} \right\|}^2} - {{\left\| {{\bf{q}}_U^{\left( l \right)}\left[ n \right] - {{\bf{q}}_k}} \right\|}^2}} \right)}}{{{\sigma ^2}\ln \left( 2 \right)\left( {{H_U}^2 + {{\left\| {{\bf{q}}_U^{\left( l \right)}\left[ n \right] - {{\bf{q}}_k}} \right\|}^2}} \right)\left( {{H_U}^2 + {{\left\| {{\bf{q}}_U^{\left( l \right)}\left[ n \right] - {{\bf{q}}_k}} \right\|}^2} + \frac{{{P_U}\left[ n \right]{\rho _{{\mathrm{com}}}}}}{{{\sigma ^2}}}} \right)}}\label{qu_Ruk}
		\end{align}
		\hrulefill	
	\end{figure*}
	\setcounter{equation}{\value{TempEqCnt}} 
	
	\section{Proposed solution}
	\label{sec:proposed solution}
	
	To solve $\mathcal{P}_{0}$, the operator ${\left[  \cdot  \right]^ + }$ is ignored because the ASR can be obtained by setting the transmit power to zero when the ASR is negative \cite{LeiH2023IoT}.
	Then, the Block Coordinate Descent (BCD) technique is utilized to decompose the original problem into multiple subproblems. Specifically, for given values of the other variables, we sequentially optimize ${\mathbf{A}}$, ${\mathbf{Y}}$,${\mathbf{P}}$, ${\mathbf{W}}$, ${{\mathbf{Q}}_{\mathrm{U}}}$, and ${{\mathbf{Q}}_{\mathrm{J}}}$.
	\color{black}
	
	\subsection{Subproblem 1: User scheduling optimization}
	
	In this subsection, for given $\left\{ {\mathbf{Y}},{{\mathbf{P}}, {\mathbf{W}}, {{\mathbf{Q}}_{\mathrm{U}}},{{\mathbf{Q}}_{\mathrm{J}}}} \right\}$, the scheduling variable ${\mathbf{A}}$ is optimized. 
	The binary constraint ${\alpha _k}\left[ n \right] \in \left\{ {0,1} \right\}$ is relaxed to $0 \leq {\alpha _k}\left[ n \right] \leq 1$ and $\mathcal{P}_{0}$ is reformulated as 
	\begin{subequations}
		\begin{align}
			\mathcal{P}_{1.1}: 	\quad&\mathop {\max }\limits_{\bf{A}} {{\hat R}_{\sec }} \label{P1.1_a}\\
			{\mathrm{s.t.}}\quad &0\le{\alpha _k}\left[ n \right] \le 1, \forall k,n				\label{P1.1_b}\\
			& \mathrm{Eqs.}\,\,  (\mathrm{\ref{P0_b}}) \,\, \mathrm{and}\,\,(\mathrm{\ref{P0_i}})										\nonumber
		\end{align}
	\end{subequations}
	where 
	${{\hat R}_{\sec }} = \frac{1}{N}\sum\limits_{n = 1}^N {\sum\limits_{k = 1}^K {{\eta }\left[ n \right]{\alpha _k}\left[ n \right]{{\tilde R}_{\sec ,k}}\left[ n \right]} }$ and 
	${{\tilde R}_{\sec ,k}}\left[ n \right] = { {{R_{Uk}}\left[ n \right] - \mathop {\max }\limits_{{{\mathbf{q}}_E} \in \mathbf{q}_E^0} {R_{UE}}\left[ n \right]} }$.
	$\mathcal{P}_{1.1}$ is a standard linear optimization problem and can be efficiently solved using CVX.

	\subsection{Subproblem 2: Optimizing the time allocation}

	In this subsection, for given $\left\{{\mathbf{A}}, {{\mathbf{P}}, {\mathbf{W}}, {{\mathbf{Q}}_{\mathrm{U}}},{{\mathbf{Q}}_{\mathrm{J}}}} \right\}$, the time allocation variable ${\mathbf{Y}}$ is optimized. 
	It can be observed that $\mathcal{P}_{1.2}$ is inherently a convex optimization problem given by 
	\begin{subequations}
		\begin{align}
			\mathcal{P}_{1.2}: 	\quad&\mathop {\max }\limits_{\bf{Y}} {{\hat R}_{\sec }} 	\label{P1.2_a}\\
			{\mathrm{s.t.}} \quad &\mathrm{Eqs.}\,\, (\mathrm{\ref{P0_i}}), (\mathrm{\ref{P0_j}}), (\mathrm{\ref{P0_k}}), \,\, \mathrm{and}\,\, (\mathrm{\ref{P0_p}}).										\nonumber
		\end{align}
	\end{subequations}
	$\mathcal{P}_{1.2}$ is a standard linear optimization problem and can be efficiently solved using CVX.
	
	\color{black}
	\subsection{Subproblem 3: Optimizing the transmit power of $U$}
	
	In this subsection, for given $\left\{ {{\mathbf{A}},{\mathbf{Y}}, {\mathbf{W}}, {{\mathbf{Q}}_{\mathrm{U}}},{{\mathbf{Q}}_{\mathrm{J}}}} \right\}$, the transmit power variable ${\mathbf{P}}$ of $U$ is optimized. Thus, $\mathcal{P}_{0}$ is reformulated as
	\color{black}
	\begin{subequations}
		\begin{align}
			\mathcal{P}_{1.3}: \quad &\mathop {\max }\limits_{\bf{P}} {{\hat R}_{\sec }}
			\label{P1.3_a}\\
			{\mathrm{s.t.}} \quad &\mathrm{Eqs.}\,\,(\mathrm{\ref{P0_d}}),(\mathrm{\ref{P0_i}}),(\mathrm{\ref{P0_j}}), \,\, \mathrm{and}\,\, (\mathrm{\ref{P0_k}}) \nonumber
		\end{align}
	\end{subequations}
	Owing to the non-convexity of both the objective function in Eq. (\ref{P1.3_a}) and the constraints given by Eqs. (\ref{P0_i}) and (\ref{P0_j}) with respective to ${\bf{P}}$, $\mathcal{P}_{1.3}$ is non-convex. 
	Utilizing Successive Convex Approximation (SCA) technology, 
	$R_{UE}\left[ n \right]$ in Eq. (\ref{P1.3_a}) and $R^{{\mathrm{sens}}}\left[ n \right]$ in Eq. (\ref{P0_j}) are approximated by their first-order Taylor expansions, which are expressed as
	\begin{align}
		{R_{UE, 1}}\left[ n \right]& = {\log _2}\left( {1 + \frac{{{h_{UE}}\left[ n \right]P_U^{\left( l \right)}\left[ n \right]}}{{{\sigma ^2} + {\mathrm{tr}}\left( {{{\bf{H}}_{JE}}\left[ n \right]{\bf{W}}\left[ n \right]} \right)}}} \right) \nonumber\\
		&+ \frac{{{h_{UE}}\left[ n \right]\left( {{P_U}\left[ n \right] - P_U^{\left( l \right)}\left[ n \right]} \right)}}{{\ln \left( 2 \right)\left( {{\sigma ^2} + {\mathrm{tr}}\left( {{{\bf{H}}_{JE}}\left[ n \right]{\bf{W}}\left[ n \right]} \right) + {h_{UE}}\left[ n \right]{P^{\left( l \right)}}\left[ n \right]} \right)}}
		\label{}
	\end{align}
	and
	\begin{align}
		R_1^{{\mathrm{sens}}}\left[ n \right] &= {\log _2}\left( {1 + \frac{{P_U^{\left( l \right)}\left[ n \right]{h_{\mathrm{sens}}}\left[ n \right]}}{{{\sigma ^2}}}} \right) + \frac{{{h_{\mathrm{sens}}}\left[ n \right]\left( {{P_U}\left[ n \right] - P_U^{\left( l \right)}\left[ n \right]} \right)}}{{\ln \left( 2 \right)\left( {{\sigma ^2} + P_U^{\left( l \right)}\left[ n \right]{h_{\mathrm{sens}}}\left[ n \right]} \right)}}
		\label{}
	\end{align}
	respectively, 
	where $P_U^{\left( l \right)}\left[ n \right]$ is a feasible point of $P_U\left[ n \right]$ at the $l$-th iteration.
	Then, $\mathcal{P}_{1.3}$ is reformulated as
	\begin{subequations}
		\begin{align}
			\mathcal{P}_{1.3{\textrm b}}: \quad &\mathop {\max }\limits_{\bf{P}} \frac{1}{N}\sum\limits_{n = 1}^N {\sum\limits_{k = 1}^K {{\eta}\left[ n \right]{\alpha _k}\left[ n \right]} } {{\tilde R}_{\sec , 1, k}}\left[ n \right]
			\label{}\\
			\text{s.t.} \quad &{R_{UJ}}\left[ n \right] \ge \mathop {\max }\limits_{{{\mathbf{q}}_E} \in \mathbf{q}_E^0} {R_1^{{\mathrm{sens}}}}\left[ n \right], \forall n \\
			&  \sum\limits_{n = 1}^N {\eta \left[ n \right]{\alpha _k}\left[ n \right]{R_{\sec ,1,k}}\left[ n \right]}  > {R _{\min }^{{\mathrm{sec}}}}, \forall k \\
			&(\mathrm{\ref{P0_d}}) \,\, \mathrm{and}\,\, (\mathrm{\ref{P0_k}}) \nonumber
		\end{align}
	\end{subequations}
	where
	
	${{\tilde R}_{\sec , 1, k}}\left[ n \right] = {R_{Uk}}\left[ n \right] - \mathop {\max }\limits_{{{\bf{q}}_E} \in {\bf{q}}_E^0} {R_{UE, 1}}\left[ n \right]$ . Thus, $\mathcal{P}_{1.3{\textrm b}}$ is a standard convex optimization problem.

	\color{black}
	\subsection{Subproblem 4: Transmit beamforming optimization} 
	
	In this subsection, for given  $\left\{ {{\bf{A}}, {\bf{Y}}, {\bf{P}}, {{\bf{Q}}_{\mathrm{U}}}, {{\bf{Q}}_{\mathrm{J}}}}\right\}$, the transmit beamforming vector ${\bf{W}}$ of $J$ is optimized. Thus,  $\mathcal{P}_{0}$ is reformulated as
	\color{black}
	\begin{subequations}
		\begin{align}
			\mathcal{P}_{1.4}: \quad &\mathop {\max }\limits_{\bf{W}} {{\hat R}_{\sec }}
			\label{P1.4_a}\\
			{\mathrm{s.t.}} \quad &\mathrm{Eqs.}\,\,(\mathrm{\ref{P0_f}})-(\mathrm{\ref{P0_i}}).\nonumber
		\end{align}
	\end{subequations}
	Based on Eq. (\ref{secRate0}), the dependence of ${R_{UE}}\left[ n \right]$ in Eqs. (\ref{P1.4_a}) and (\ref{P0_i}) on ${\bf W}$ renders $\mathcal{P}_{1.4}$ non-convex.  
	We approximate ${R_{UE}}\left[ n \right]$ using the first-order Taylor expansion with respect to ${\mathbf{W}}$, which is expressed as Eq. (\ref{RUEin13}), shown at the top of this  page.  
	Then, $\mathcal{P}_{1.4}$ is reformulated as \setcounter{equation}{24}
	\begin{subequations}
		\begin{align}
			\mathcal{P}_{1.4{\textrm b}}: \quad &\mathop {{\mathrm{max}}}\limits_{\bf{W}} \frac{1}{N}\sum\limits_{n = 1}^N {\sum\limits_{k = 1}^K {{\eta}\left[ n \right]{\alpha _k}\left[ n \right]{{\tilde R}_{\sec, 2, k}}} } \left[ n \right] 	\label{P1.3b_a}\\
			\text{s.t.} \quad &  \sum\limits_{n = 1}^N {\eta \left[ n \right]{\alpha _k}\left[ n \right]{R_{\sec ,2,k}}\left[ n \right]}  > {R _{\min }^{{\mathrm{sec}}}}, \forall k \\
			&\mathrm{Eqs.}\,\, (\mathrm{\ref{P0_f}}) \,\, \mathrm{and}\,\, (\mathrm{\ref{P0_g}}) \nonumber
		\end{align}
	\end{subequations}
	where 
	
	${{\tilde R}_{\sec, 2, k}}\left[ n \right] = {R_{Uk}}\left[ n \right] - \mathop {\max }\limits_{{{\bf{q}}_E} \in {\bf{q}}_E^0} {R_{UE, 2}}\left[ n \right]$. 
	By ignoring rank-1 constraint Eq. (\ref{P0_h}), ${\mathcal{P}_{1.4{\textrm b}}}$ is a Semi-Definite Relaxation (SDR) optimization problem that can be efficiently solved using CVX\footnote{
		The rank-1 constraint can be handled by Gaussian randomization \cite{ChengG2025TCOM}, the method based on a penalty function \cite{DanQ2025TCCN}, or the iterative method proposed in \cite{DanQ2025TVT}.
	}. 
	
	\color{black}
	\subsection{Subproblem 5: Trajectory optimization of U}
	
	In this subsection, for given  $\{{\bf{A}},{\bf{Y}}, {\bf{P}}, {\bf{W}}, {{\bf{Q}}_{\mathrm{J}}}\}$, the flight trajectory ${{\bf{Q}}_{\mathrm{U}}}$ of $U$ is optimized. $\mathcal{P}_{0}$ is reformulated as
	\color{black}
	\begin{subequations}
		\begin{align}
			\mathcal{P}_{1.5}: \quad &\mathop {{\mathrm{max}}}\limits_{{{\bf{Q}}_{\mathrm{U}}}} {{\hat R}_{\sec }}
			\label{P1.5_a}\\
			{\mathrm{s.t.}} \quad &\mathrm{Eqs.}\,\,(\mathrm{\ref{P0_E}}),(\mathrm{\ref{P0_i}})-(\mathrm{\ref{P0_o}}) \nonumber
		\end{align}
	\end{subequations}
	It should be noted that the objective function in 
	Eq. (\ref{P1.5_a}), and the constraints given by Eqs. (\ref{P0_E}), and (\ref{P0_j}) are non-convex with respect to ${{\bf{Q}}_{\mathrm{U}}}$, thereby resulting in a non-convex optimization problem.
	
	First, using first-order Taylor expansion with respect to ${{\mathbf{Q}}_{\mathrm{U}}}$, 
	$R_{Uk}\left[ n \right]$ is approximated as Eq. (\ref{qu_Ruk}), shown on the top of this page, 
	where ${{\bf{q}}_U^{\left( l \right)}\left[ n \right]}$ is a feasible point of ${{{\bf{q}}_U}\left[ n \right]}$ at the $l$-th iteration.
	Then, by introducing a slack variable ${\xi \left[ n \right]}$, $R_{UE}\left[ n \right]$ and $R^{{\mathrm{sens}}}\left[ n \right]$ are expressed as \setcounter{equation}{27}
	
	\begin{align}
		{R_{UE, 3}}\left[ n \right] &= {\log _2}\left( {1 + \frac{{{P_U}\left[ n \right]{\rho _{{\mathrm{com}}}}}}{{\left( {{\sigma ^2} + {\mathrm{tr}}\left( {{{\bf{H}}_{JE}}\left[ n \right]{\bf{W}}\left[ n \right]} \right)} \right)\xi \left[ n \right]}}} \right) \label{quRUE} 
	\end{align}
	\color{black}
	and
	\begin{align}
		R_2^{{\mathrm{sens}}}\left[ n \right] = {\log _2}\left( {1 + \frac{{{P_U}\left[ n \right]{\rho _{{\mathrm{sens}}}}}}{{{\sigma ^2}\xi {{\left[ n \right]}^2}}}} \right)
		\label{}
	\end{align}
	respectively, 
	subjective to the following constraint
	\begin{align}
		\xi \left[ n \right]&  \le {\left\| {{\bf{q}}_U^{\left( l \right)}\left[ n \right] - {{\bf{q}}_E}} \right\|^2} + 2{\left( {{\bf{q}}_U^{\left( l \right)}\left[ n \right] - {{\bf{q}}_E}} \right)^T}\left( {{{\bf{q}}_U}\left[ n \right] - {\bf{q}}_U^{\left( l \right)}\left[ n \right]} \right)
		\label{xiyueshusca}
	\end{align}
	
	To handle the term $R_{UJ}\left[ n \right]$ in Eq. (\ref{P0_j}), we approximate $R_{UJ}\left[ n \right]$ as Eq. (\ref{quUJsca}), shown at the top of the next page.
	\begin{figure*}[ht]
		\begin{align}
			{R_{UJ, 1}}\left[ n \right] &= {\log _2}\left( {1 + \frac{{{P_U}\left[ n \right]{\rho _{\mathrm{com}}}}}{{{\sigma ^2}\left( {{{\left\| {{H_U} - {H_J}} \right\|}^2} + {{\left\| {{\bf{q}}_U^{\left( l \right)}\left[ n \right] - {{\bf{q}}_J}\left[ n \right]} \right\|}^2}} \right)}}} \right)  \notag\\
			&- \frac{{{P_U}\left[ n \right]{\rho _{\mathrm{com}}}\left( {{{\left\| {{{\bf{q}}_U}\left[ n \right] - {{\bf{q}}_J}\left[ n \right]} \right\|}^2} - {{\left\| {{\bf{q}}_U^{\left( l \right)}\left[ n \right] - {{\bf{q}}_J}\left[ n \right]} \right\|}^2}} \right)}}{{{\sigma ^2}\ln \left( 2 \right)\left( {{{\left\| {{H_U} - {H_J}} \right\|}^2} + {{\left\| {{\bf{q}}_U^{\left( l \right)}\left[ n \right] - {{\bf{q}}_J}\left[ n \right]} \right\|}^2}} \right)\left( {{{\left\| {{H_U} - {H_J}} \right\|}^2} + {{\left\| {{\bf{q}}_U^{\left( l \right)}\left[ n \right] - {{\bf{q}}_J}\left[ n \right]} \right\|}^2} + \frac{{{P_U}\left[ n \right]{\rho _{\mathrm{com}}}}}{{{\sigma ^2}}}} \right)}} 
			\label{quUJsca}
		\end{align}
		\hrulefill
	\end{figure*}
	\setcounter{equation}{38} 
	\begin{figure*}[h]
		\begin{align}
			{R_{UJ,2}}\left[ n \right]&= {\log _2}\left( {1 + \frac{{{P_U}\left[ n \right]{\rho _{\mathrm{com}}}}}{{{\sigma ^2}\left( {{{\left\| {{H_U} - {H_J}} \right\|}^2} + {{\left\| {{{\bf{q}}_U}\left[ n \right] - {\bf{q}}_J^{\left( l \right)}\left[ n \right]} \right\|}^2}} \right)}}} \right) \notag\\
			&- \frac{{{P_U}\left[ n \right]{\rho _{\mathrm{com}}}\left( {{{\left\| {{{\bf{q}}_U}\left[ n \right] - {{\bf{q}}_J}\left[ n \right]} \right\|}^2} - {{\left\| {{{\bf{q}}_U}\left[ n \right] - {\bf{q}}_J^{\left( l \right)}\left[ n \right]} \right\|}^2}} \right)}}{{{\sigma ^2}\ln \left( 2 \right)\left( {{{\left\| {{H_U} - {H_J}} \right\|}^2} + {{\left\| {{{\bf{q}}_U}\left[ n \right] - {\bf{q}}_J^{\left( l \right)}\left[ n \right]} \right\|}^2}} \right)\left( {{{\left\| {{H_U} - {H_J}} \right\|}^2} + {{\left\| {{{\bf{q}}_U}\left[ n \right] - {\bf{q}}_J^{\left( l \right)}\left[ n \right]} \right\|}^2} + \frac{{{P_U}\left[ n \right]{\rho _{\mathrm{com}}}}}{{{\sigma ^2}}}} \right)}}
			\label{RUJSCA}
		\end{align}
		\hrulefill
	\end{figure*} 
	Thus, Eq. (\ref{P0_i}) is reformulated as \setcounter{equation}{31} 
	\begin{align}
		{R_{UJ, 1}}\left[ n \right] \ge \mathop {\max }\limits_{{{\mathbf{q}}_E} \in \mathbf{q}_E^0} {R_2^{{\mathrm{sens}}}}\left[ n \right], \forall n
		\label{UJcomm}
	\end{align}
	
	Next, following the method in \cite{LeiH2024TCCN} and introducing a slack variable ${{\lambda _U}\left[ n \right]}$, Eq. (\ref{P0_E}) is reformulated as
	\begin{align}
		\frac{1}{N}\sum_{n = 1}^N &\left[ P_0 \left( 1 + \frac{3\lVert \mathbf{v}_U\left[ n \right] \rVert^2}{U_{\text{tip}}^2} \right) + \frac{1}{2}d_0\rho sA\lVert \mathbf{v}_U\left[ n \right] \rVert^3 + P_1\lambda_U\left[ n \right] \right] \leq P_{U,\text{ave}}^{\text{fly}}
		\label{quflypower}
	\end{align}
	and
	\begin{align}
		\frac{1}{{{\lambda _U}{{\left[ n \right]}^2}}} &\le \lambda _U^{\left( l \right)}{\left[ n \right]^2} + 2\lambda _U^{\left( l \right)}\left[ n \right]\left( {{\lambda _U}\left[ n \right] - \lambda _U^{\left( l \right)}\left[ n \right]} \right) \notag\\
		&+ \frac{\left\|{\mathbf{v}}^{{\left( l \right)}}_{U}\left[ n \right]\right\|^2}{v_0^2} + \frac{2}{{v_0^2}}{\left( {{\bf{v}}_U^{\left( l \right)}\left[ n \right]} \right)^T}\left( {{{\bf{v}}_U}\left[ n \right] - {\bf{v}}_U^{\left( l \right)}\left[ n \right]} \right)
		\label{lambdasongchisca}
	\end{align}
	respectively, 
	where 
	$\lambda_U^{\left( l \right)}\left[ n \right]$ is a feasible point of ${\lambda_U}\left[ n \right]$ at the $l$-th iteration.

	Finally, $\mathcal{P}_{1.5}$ is approximated as
	\begin{subequations}
		\begin{align}
			\mathcal{P}_{1.5{\textrm b}}: \quad &\mathop {\max }\limits_{{{\bf{Q}}_{\mathrm{U}}},{{\bf{v}}_{\mathrm{U}}},{\lambda _U},\xi } \frac{1}{N}\sum\limits_{n = 1}^N {\sum\limits_{k = 1}^K {{\eta }\left[ n \right]} {\alpha _k}\left[ n \right]{{\tilde R}_{\sec, 3, k}}\left[ n \right]} 
			\label{P4.2_a}\\
			\text{s.t.} \quad &  \sum\limits_{n = 1}^N {\eta \left[ n \right]{\alpha _k}\left[ n \right]{R_{\sec ,3,k}}\left[ n \right]}  > {R _{\min }^{{\mathrm{sec}}}}, \forall k \\
			&\mathrm{Eqs.}\,\,(\ref{xiyueshusca}), (\ref{UJcomm}), (\ref{quflypower}), (\ref{lambdasongchisca}), (\mathrm{\ref{P0_k}})-(\mathrm{\ref{P0_o}}) \nonumber
		\end{align}
	\end{subequations}
	where 
	${{\tilde R}_{\sec, 3, k}}\left[ n \right] = {R_{Uk, 1}}\left[ n \right] - \mathop {\max }\limits_{{{\bf{q}}_E} \in {\bf{q}}_E^0} {R_{UE, 3}}\left[ n \right]$. Then,$\mathcal{P}_{1.5{\textrm b}}$ is a convex optimization problem.
	
	\subsection{Subproblem 6: Optimization of the flight trajectory of $J$}

	In this subsection, for given  $\left\{ {{\bf{A}}, {\bf{Y}}, {\bf{P}}, {\bf{W}}, {{\bf{Q}}_{\mathrm{U}}}} \right\}$, the trajectory ${{\bf{Q}}_{\mathrm{J}}}$ of $J$ is optimized.  $\mathcal{P}_{0}$ is reformulated as
	\color{black}
	\begin{subequations}
		\begin{align}
			\mathcal{P}_{1.6}: \quad &\mathop {\max }\limits_{{{\bf{Q}}_{\mathrm{J}}}} {{\hat R}_{\sec }}
			\label{P1.6_a}\\
			{\mathrm{s.t.}} \quad &\mathrm{Eqs.}\,\,(\mathrm{\ref{P0_E}}),(\mathrm{\ref{P0_i}})-(\mathrm{\ref{P0_o}}) \nonumber
		\end{align}
	\end{subequations}
	In $\mathcal{P}_{1.6}$, Eqs.  (\ref{P0_E}), (\ref{P0_j}), and (\ref{P1.6_a}) are all non-convex with respective to ${{{\bf{Q}}_{\mathrm{J}}}}$. Therefore, the resulting problem is non-convex.
	
	By introducing a new slack variable $\psi \left[ n \right]$ and defining 
	${{\bf{A}}_E}\left[ n \right] = {{\bf{a}}_E}\left[ n \right]{{\bf{a}}_E}{\left[ n \right]^H}$, 
	$R_{UE}\left[ n \right]$ in Eq. (\ref{P1.6_a}) is rewritten as\footnote{
		It is worth highlighting that the steering vector ${{\mathbf{a}}_E}\left[ n \right]$ depends on the trajectory of $J$  and exhibits strong nonlinearity, further compounding the difficulty of solving $\mathcal{P}_{1.6}$. To tackle this non-convexity, we employ a trust-region SCA framework to approximate $\mathbf{a}_{E}\left[ n \right]$ within $\mathbf{h}_{JE}\left[ n \right]$, following the methods in the ISAC literature \cite{WangY2026IoT, JiangL2026IoT, LyuZ2023TWC}.
	}
	\begin{align}
		{R_{UE, 4}}\left[ n \right] = {\log _2}\left( {1 + \frac{{{P_U}\left[ n \right]{h_{UE}}\left[ n \right]}}{{{\sigma ^2} + \frac{{{\rho _{{\mathrm{com}}}}{\mathrm{tr}}\left( {{{\bf{A}}_E}\left[ n \right]{\bf{W}}\left[ n \right]} \right)}}{{\psi \left[ n \right]}}}}} \right)
		\label{}
	\end{align}
	subjective to the following constraint
	\begin{align}
		\psi \left[ n \right] & \le \left\| {\bf{q}}_J^{\left( l \right)}\left[ n \right] - {{\bf{q}}_E} \right\| ^2 \nonumber \\
		&+ 2{({\bf{q}}_J^{\left( l \right)}\left[ n \right] - {{\bf{q}}_E})^T}({{\bf{q}}_J}\left[ n \right] - {\bf{q}}_J^{\left( l \right)}\left[ n \right]) + {H_J}^2
		\label{psiyueshusca}
	\end{align}
	where ${{\bf{q}}_J^{\left( l \right)}\left[ n \right]}$ is a feasible point of ${{{\bf{q}}_J}\left[ n \right]}$ at the $l$-th iteration.
	
	Next,  $R_{UJ}\left[ n \right]$ in Eq. (\ref{P0_j}) is replaced by its first-order Taylor expansion with respect to ${{\bf{Q}}_{\mathrm{J}}}$ of $J$, which is expressed as Eq. (\ref{RUJSCA}), shown at the top of this page.
	Thus, Eq. (\ref{P0_j}) is rewritten as\setcounter{equation}{39} 
	\begin{align}
		{R_{UJ, 2}}\left[ n \right] \ge \mathop {\max }\limits_{{{\mathbf{q}}_E} \in \mathbf{q}_E^0} {R^{{\mathrm{sens}}}}\left[ n \right], \forall n
		\label{qjUJcomm}
	\end{align}
	
	Finally, following the same method as in Eq. (\ref{quflypower}), we introduce a slack variable ${{\lambda _J}\left[ n \right]}$, Eq. (\ref{P0_E}) is rewritten as
	\begin{align}
		\frac{1}{N}\sum_{n = 1}^N &\left[ P_0 \left( 1 + \frac{3\left\| {{{\bf{v}}_J}\left[ n \right]} \right\|^2}{U_{\text{tip}}^2} \right) \notag \right. \\
		&\left. + \frac{1}{2}d_0\rho sA\left\| {{{\bf{v}}_J}\left[ n \right]} \right\|^3 + P_1\lambda_J\left[ n \right] \right] \leq P_{J,\text{ave}}^{\text{fly}}	
		\label{qjflypower}
	\end{align}
	and
	\begin{align}
		\frac{1}{{{\lambda _J}{{\left[ n \right]}^2}}} &\le \lambda _J^{\left( l \right)}{\left[ n \right]^2} + 2\lambda _J^{\left( l \right)}\left[ n \right]\left( {{\lambda _J}\left[ n \right] - \lambda _J^{\left( l \right)}\left[ n \right]} \right) \notag\\
		&+ \frac{\left\|{\mathbf{v}}^{{\left( l \right)}}_{J}\left[ n \right]\right\|^2}{v_0^2} + \frac{2}{{v_0^2}}{\left( {{\bf{v}}_J^{\left( l \right)}\left[ n \right]} \right)^T}\left( {{{\bf{v}}_J}\left[ n \right] - {\bf{v}}_J^{\left( l \right)}\left[ n \right]} \right)
		\label{qjflypowersca}
	\end{align}
	respectively, 
	where 
	$\lambda_J^{\left( l \right)}\left[ n \right]$ and ${{\bf{v}}_J^{\left( l \right)}\left[ n \right]}$ are feasible points of ${\lambda_J}\left[ n \right]$  and ${{{\bf{v}}_J}\left[ n \right]}$ at the $l$-th iteration, respectively.
	Thus, $\mathcal{P}_{1.5}$ is reformulated as
	\begin{subequations}
		\begin{align}
			\mathcal{P}_{1.6{\textrm b}}: \quad &\mathop {{\mathrm{max}}}\limits_{{{\bf{Q}}_{\mathrm{J}}},\psi ,{\lambda _{\mathrm{J}}}} \frac{1}{N}\sum\limits_{n = 1}^N {\sum\limits_{k = 1}^K {{\eta}\left[ n \right]} {\alpha _k}\left[ n \right]{{\tilde R}_{\sec, 4, k}}\left[ n \right]} 		\label{P5.2_a}\\
			\text{s.t.} \quad &  \sum\limits_{n = 1}^N {\eta \left[ n \right]{\alpha _k}\left[ n \right]{R_{\sec ,4,k}}\left[ n \right]}  > {R _{\min }^{{\mathrm{sec}}}}, \forall k \\
			&\mathrm{Eqs.}\,\,  (\mathrm{\ref{P0_k}})-(\mathrm{\ref{P0_o}}), (\ref{psiyueshusca}), (\ref{qjUJcomm}), (\ref{qjflypower}), (\ref{qjflypowersca}), \nonumber
		\end{align}
	\end{subequations}
	where ${{\tilde R}_{\sec, 4, k}}\left[ n \right] = {R_{Uk, 1}}\left[ n \right] - \mathop {\max }\limits_{{{\bf{q}}_E} \in {\bf{q}}_E^0} {R_{UE, 4}}\left[ n \right]$.
	$\mathcal{P}_{1.6{\textrm b}}$ is a convex optimization problem and can be solved using CVX.

	{The overall iterative procedure for solving problem $\mathcal{P}_{0}$ is summarized in \textbf{Algorithm \ref{algorithm1}}, where $R\left( {{{\mathbf{A}}^{\left( l \right)}},{{\mathbf{Y}}^{\left( l \right)}},{{\mathbf{P}}^{\left( l \right)}},{{\mathbf{W}}^{\left( l \right)}},{{\mathbf{Q_U}}^{\left( l \right)}},{{\mathbf{Q_J}}^{\left( l \right)}}} \right)$ denotes the objective value of the original problem $\mathcal{P}_{0}$ at the $l$-th iteration, $\varepsilon$ and ${l_{\max }}$ denote the convergence tolerance and the maximum number of iterations, respectively.}
	
	\begin{algorithm}[t]
		
		\small
		\caption{Iterative Procedure of $\mathcal{P}_{0}$}
		\label{algorithm1}
		\KwIn{
			Initialize feasible points
		}
		\Do{$R\left( {{{\mathbf{A}}^{\left( {l + 1} \right)}},{{\mathbf{Y}}^{\left( {l + 1} \right)}},{{\mathbf{P}}^{\left( {l + 1} \right)}},{{\mathbf{W}}^{\left( {l + 1} \right)}},{{\mathbf{Q_U}}^{\left( {l + 1} \right)}},{{\mathbf{Q_J}}^{\left( {l + 1} \right)}}} \right) - R\left( {{{\mathbf{A}}^{\left( l \right)}},{{\mathbf{Y}}^{\left( l \right)}},{{\mathbf{P}}^{\left( l \right)}},{{\mathbf{W}}^{\left( l \right)}},{{\mathbf{Q_U}}^{\left( l \right)}},{{\mathbf{Q_J}}^{\left( l \right)}}} \right) \succ \varepsilon$ \& $l \le {l_{\max }}$}
		{   1. Obtain ${\mathbf A}^{\left( {l + 1} \right)}$ by solving ($\mathcal{P}_{1.1}$) for given ${{\mathbf{Y}}^{\left( l \right)}}$, ${{\mathbf{P}}^{\left( l \right)}}$, ${{\mathbf{W}}^{\left( l \right)}}$, ${{\mathbf{Q_U}}^{\left( l \right)}}$, and ${{\mathbf{Q_J}}^{\left( l \right)}}$;\\
			2. Obtain ${\mathbf Y}^{\left( {l + 1} \right)}$ by solving ($\mathcal{P}_{1.2}$) for given ${{\mathbf{A}}^{\left( {l + 1} \right)}}$, ${{\mathbf{P}}^{\left( l \right)}}$, ${{\mathbf{W}}^{\left( l \right)}}$, ${{\mathbf{Q_U}}^{\left( l \right)}}$, and ${{\mathbf{Q_J}}^{\left( l \right)}}$;\\
			3. Obtain ${{{\mathbf{P}}^{\left( {l + 1} \right)}}}$ by solving ($\mathcal{P}_{1.3{\textrm b}}$) for given ${{\mathbf{A}}^{\left( {l + 1} \right)}}$, ${{\mathbf{Y}}^{\left( {l + 1} \right)}}$,${{\mathbf{W}}^{\left( l \right)}}$, ${{\mathbf{Q_U}}^{\left( l \right)}}$, and ${{\mathbf{Q_J}}^{\left( l \right)}}$;\\
			4. Obtain ${{{\mathbf{W}}^{\left( {l + 1} \right)}}}$ by solving ($\mathcal{P}_{1.4{\textrm b}}$) for given ${{\mathbf{A}}^{\left( {l + 1} \right)}}$, ${{\mathbf{Y}}^{\left( {l + 1} \right)}}$,${{\mathbf{P}}^{\left( {l + 1} \right)}}$, ${{\mathbf{Q_U}}^{\left( l \right)}}$, and ${{\mathbf{Q_J}}^{\left( l \right)}}$;\\
			5. Obtain ${{{\mathbf{Q_U}}^{\left( {l + 1} \right)}}}$ by solving ($\mathcal{P}_{1.5{\textrm b}}$) for given ${{\mathbf{A}}^{\left( {l + 1} \right)}}$,${{\mathbf{Y}}^{\left( {l + 1} \right)}}$, ${{\mathbf{P}}^{\left( {l + 1} \right)}}$, ${{\mathbf{W}}^{\left( {l + 1} \right)}}$, and ${{\mathbf{Q_J}}^{\left( l \right)}}$;\\
			6. Obtain ${{{\mathbf{Q_J}}^{\left( {l + 1} \right)}}}$ by solving ($\mathcal{P}_{1.6{\textrm b}}$) for given ${{\mathbf{A}}^{\left( {l + 1} \right)}}$,${{\mathbf{Y}}^{\left( {l + 1} \right)}}$, ${{\mathbf{P}}^{\left( {l + 1} \right)}}$, ${{\mathbf{W}}^{\left( {l + 1} \right)}}$, and ${{\mathbf{Q_U}}^{\left( {l + 1} \right)}}$;\\
			7. $l = l + 1$;
		}
		\KwOut{$R\left( {{{\mathbf{A}}^{\left( l \right)}},{{\mathbf{Y}}^{\left( l \right)}},{{\mathbf{P}}^{\left( l \right)}},{{\mathbf{W}}^{\left( l \right)}},{{\mathbf{Q_U}}^{\left( l \right)}},{{\mathbf{Q_J}}^{\left( l \right)}}} \right)$}
	\end{algorithm}
	
	\begin{figure*}[t]
		\centering
		\subfigure[]{
			\label{fig03a}
			\includegraphics[width = \figwidThree, height = \figheiThree]{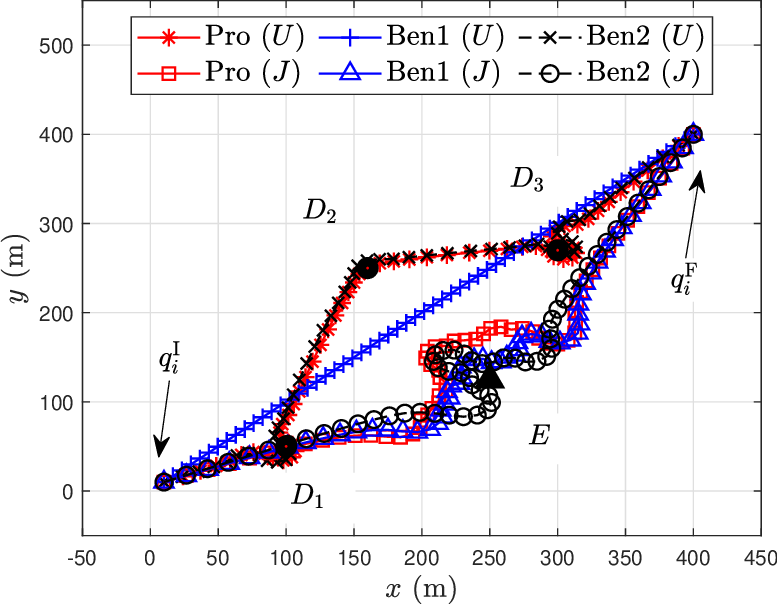}}
		\subfigure[]{
			\label{fig03b}
			\includegraphics[width = \figwidThree, height = \figheiThree]{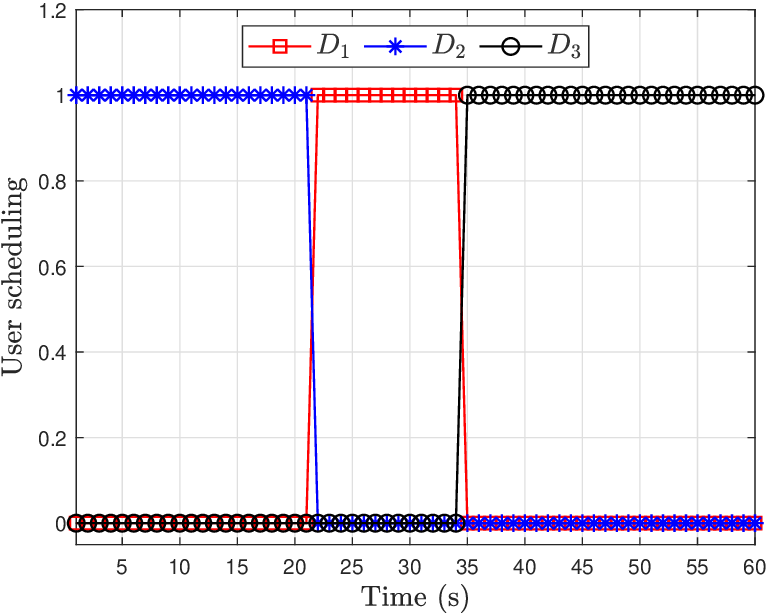}}
		\subfigure[]{
			\label{fig03c}
			\includegraphics[width = \figwidThree, height = \figheiThree]{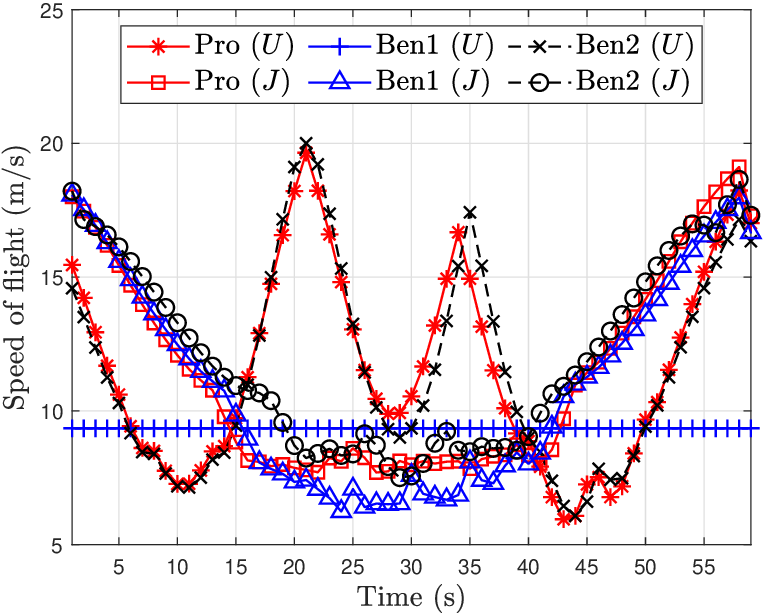}}
		\subfigure[]{
			\label{fig03d}
			\includegraphics[width = \figwidThree, height = \figheiThree]{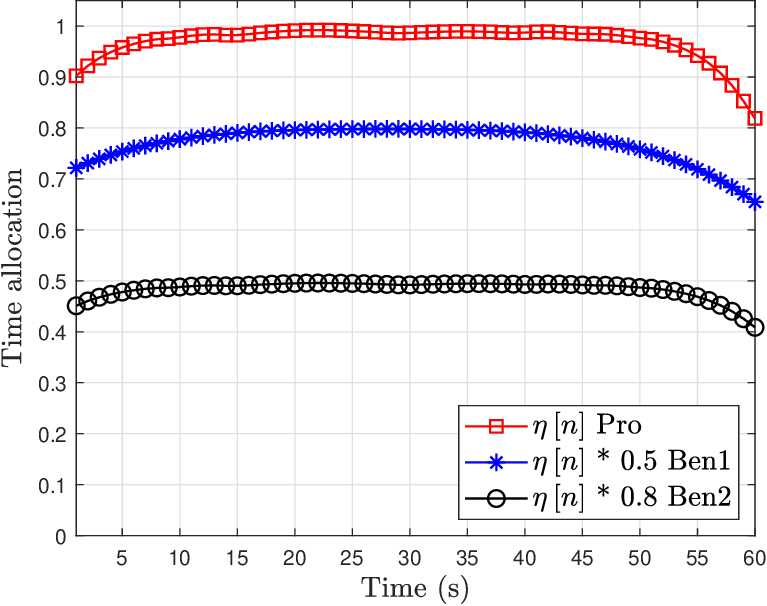}}
		\subfigure[]{
			\label{fig03e}
			\includegraphics[width = \figwidThree, height = \figheiThree]{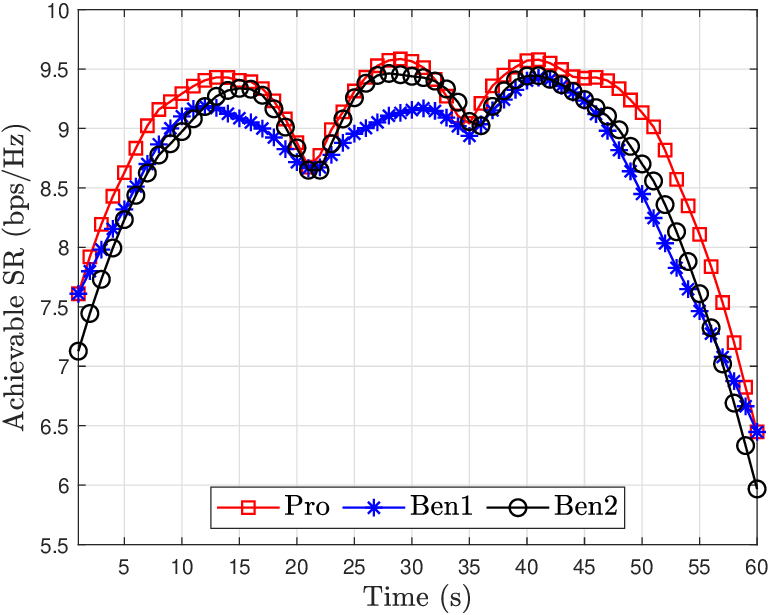}}
		\subfigure[]{
			\label{fig03f}
			\includegraphics[width = \figwidThree, height = \figheiThree]{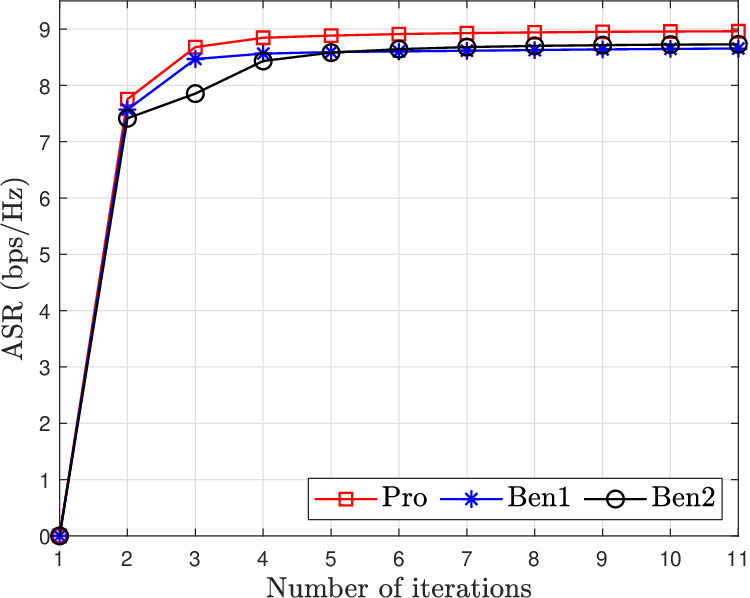}}
		\caption{Scenario 1 wherein $K = 3$ users are distributed on two sides of the initial trajectory with $N = 60$ and $P_{{\mathrm{ave}}}^{{\mathrm{fly}}} = 140$. (a) Optimal trajectories. (b) User scheduling. (c) Speed of flight. (d) Time allocation. (e) Achievable SR. (f) ASR versus the number of iterations.  }
		\label{fig03}
	\end{figure*}
	
	\begin{figure*}[t]
		\centering
		\subfigure[]{
			\label{fig04a}
			\includegraphics[width = \figwidThree, height = \figheiThree]{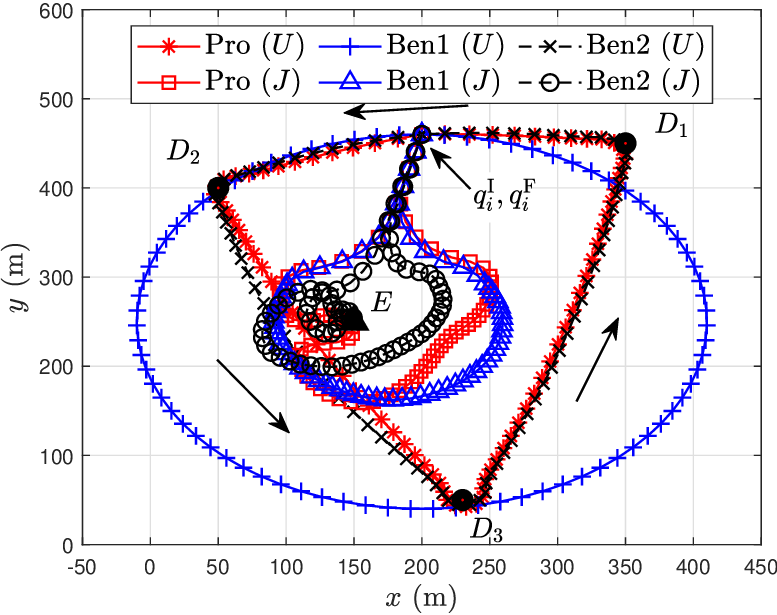}}
		\subfigure[]{
			\label{fig04b}
			\includegraphics[width = \figwidThree, height = \figheiThree]{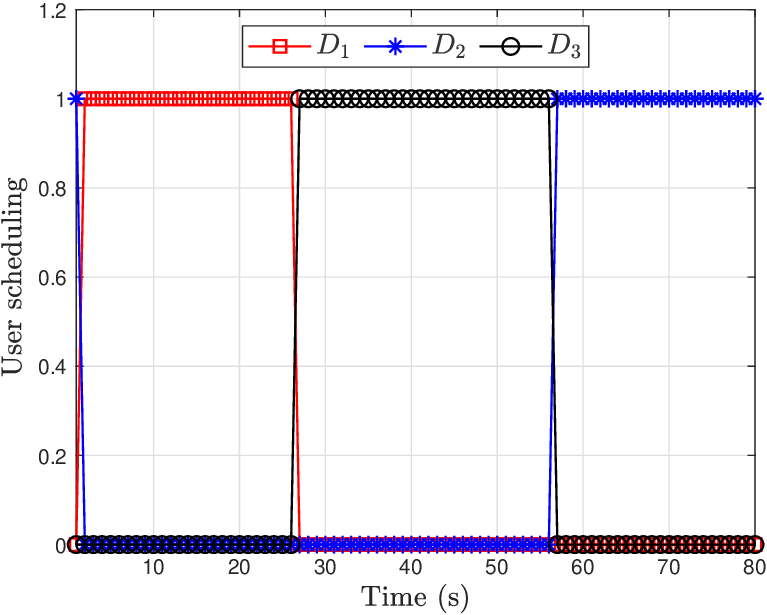}}
		\subfigure[]{
			\label{fig04c}
			\includegraphics[width = \figwidThree, height = \figheiThree]{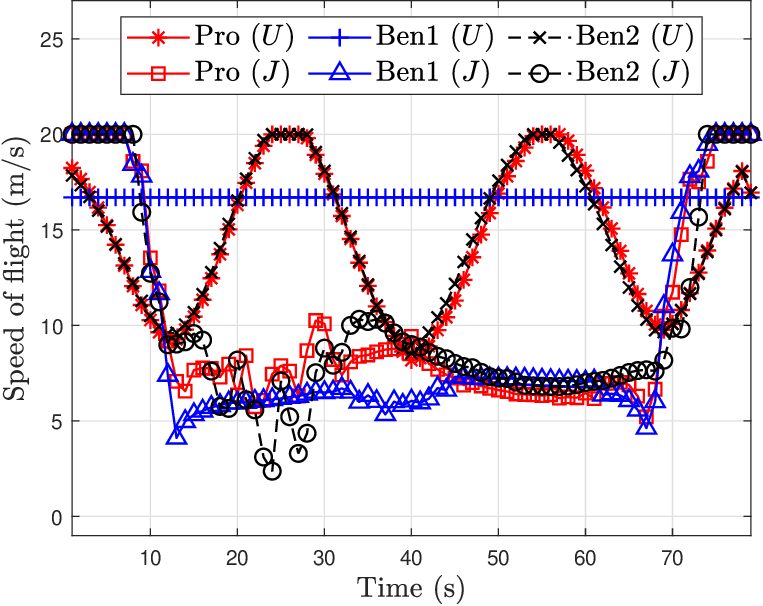}}
		\subfigure[]{
			\label{fig04d}
			\includegraphics[width = \figwidThree, height = \figheiThree]{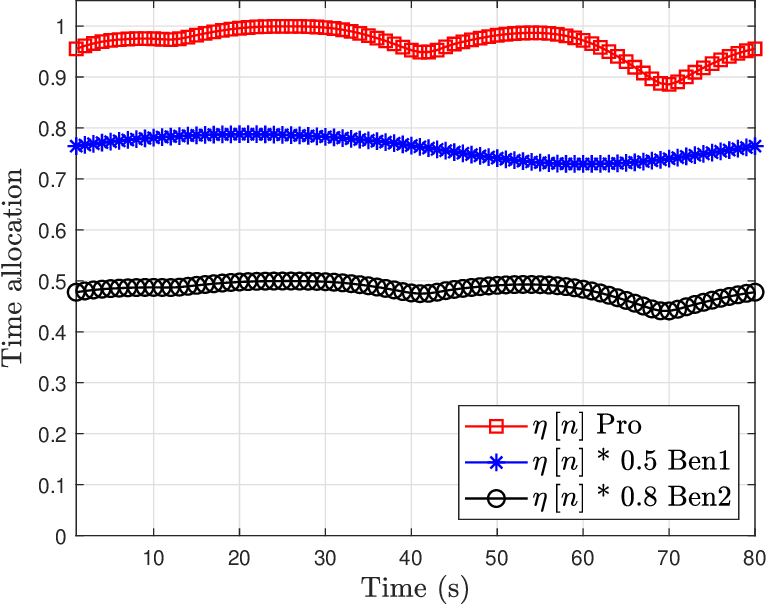}}
		\subfigure[]{
			\label{fig04e}
			\includegraphics[width = \figwidThree, height = \figheiThree]{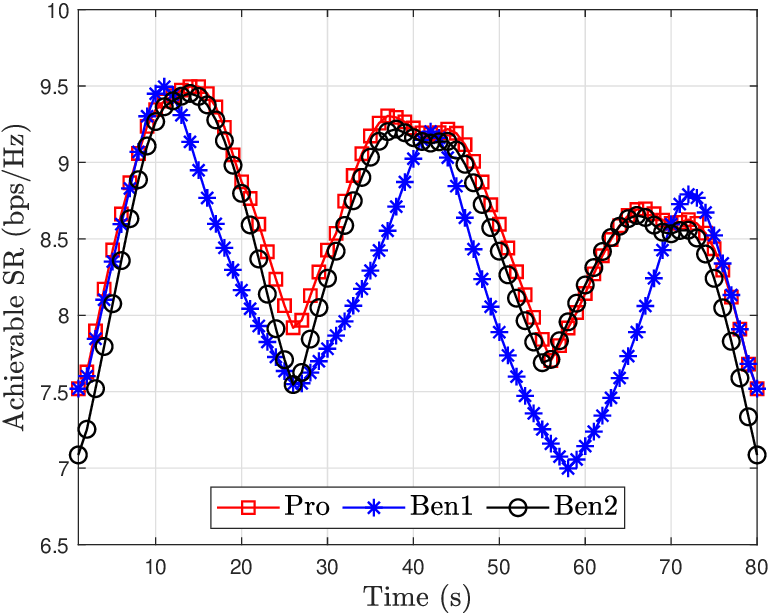}}
		\subfigure[]{
			\label{fig04f}
			\includegraphics[width = \figwidThree, height = \figheiThree]{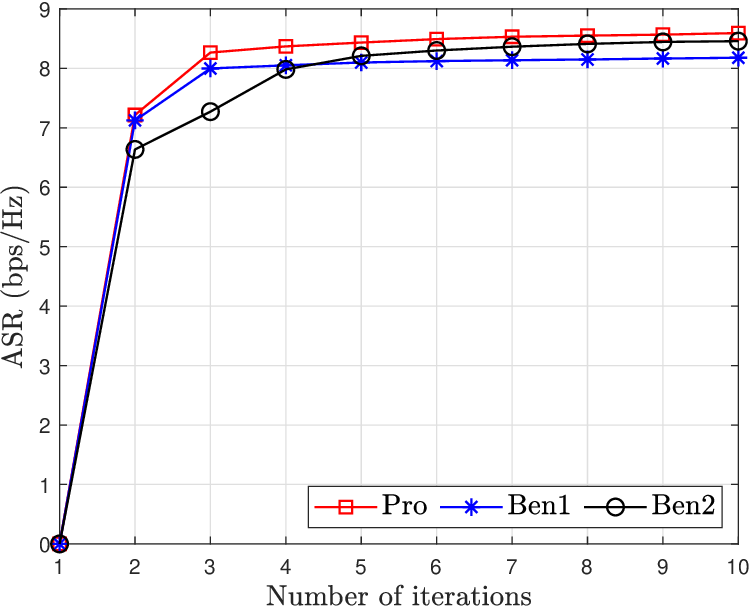}}
		\caption{Scenario 2 wherein $K = 3$ users are distributed around $E$ with $N = 80$ and $P_{{\mathrm{ave}}}^{{\mathrm{fly}}} = 150$. (a) Optimal trajectories. (b) User scheduling. (c) Speed of flight.  (d) Time allocation. (e) Achievable SR. (f) ASR versus the number of iterations.}
		\label{fig04}
	\end{figure*}
	
	{
		\subsection{Complexity analysis}
		\label{sec:Algorithm}
		
		This section analyzes the computational complexity of the proposed iterative algorithm. Based on the BCD iterative framework, the original non-convex, multi-variable, coupled joint optimization problem is decomposed into five efficiently solvable convex subproblems.
		
		$\mathcal{P}_{1.1}$ and $\mathcal{P}_{1.2}$ are formulated as standard linear programming problems. These problems contain only $K N$-dimensional scheduling coefficient optimization variables, as well as boundary and global summation constraints with corresponding dimensions. This linear programming problem is solved by the interior-point method. The final computational complexity of this sub-problem is $\mathcal{O}\left((KN)^{3.5}\log(1/\varepsilon)\right)$.
		
		$\mathcal{P}_{1.3{\textrm b}}$ is the transmit power optimization of $U$. By adopting the first-order Taylor convex approximation, the original non-convex communication and sensing rate expressions are transformed into a standard convex cone programming problem. The optimization variables of this subproblem are $N$-dimensional time-slot transmit power variables. The dimensions of variables and constraints are jointly determined by the number of users $\left( {K} \right)$, the number of discrete sampling points of $E$ $\left( {M} \right)$, and the number of time slots $\left( {N} \right)$. The corresponding computational complexity of this cone programming problem solved by the interior-point method is $\mathcal{O}\left(((K + M)N)^{3.5}\log(1/\varepsilon)\right)$.
		
		$\mathcal{P}_{1.4{\textrm b}}$ corresponds to the beamforming matrix optimization of $J$. By employing the  SDR technique, the non-convex rank constraint is relaxed into a positive semi-definite constraint, thereby formulating a standard semi-definite programming problem. The solution dimension is determined by the antenna size $\left( {N_t} \right)$ and the number of time slots $\left( {N} \right)$, yielding a complexity of $\mathcal{O}\left(\left(N_t^2N\right)^{3.5}\log(1/\varepsilon)\right)$. 
		
		$\mathcal{P}_{1.5{\textrm b}}$ and $\mathcal{P}_{1.6{\textrm b}}$ represent the trajectory optimization of UAVs. Both trajectory optimization subproblems convexify the non-convex distance and rate constraints via the SCA method and are finally transformed into convex cone programming problems. Their corresponding computational complexities are  $\mathcal{O}\left((N_tN)^{3.5}\log(1/\varepsilon)\right)$ and $\mathcal{O}\left(((K + N_t)N)^{3.5}\log(1/\varepsilon)\right)$, respectively.
		Accordingly, combined with the outer iteration number, the overall computational complexity of the proposed algorithm is summarized as $\mathcal{O}\left({l_{\max }}\left((N_t^2N)^{3.5}+((K + M)N)^{3.5}\right)\log\frac{1}{\varepsilon}\right)$.

		{\color{black}
			\section{Numerical results and analysis}
			\label{sec:Simulation Results}

			\begin{table}[t]
				\centering
				\begin{minipage}{0.4\textwidth}
					\captionsetup{justification=raggedright,singlelinecheck=false}
					\caption{Simulation Parameters}
					\label{table3}
					\resizebox{\linewidth}{!}{
						\begin{tabular}{c c c c}
							\toprule
							\textbf{Notation}   	& \textbf{Value}  & \textbf{Notation}   	& \textbf{Value}  \\
							\midrule
							${H_U}$	&  $60$ m  &	${H_J}$	&  50 m  \\
							${\delta }$		& $1 $ s   &   $P_{{\mathrm{ave}}}^{{\mathrm{fly}}}$ &   $150$ W \\
							${N_t}$ 			& $4$    	& ${P_0}$ &  $79.86$ W\\
							${\sigma^2}$ 	& $- 90$ dBm  & ${P_i}$  & $88.63$ W \\
							$A$ & $0.503$ ${{\mathrm m}^2}$ &  ${U_{tip}}$& $120$ m/s\\
							${P_{\max }}$ 	& 5 W &  ${v_0}$&  $4.03$ m/s \\
							${{a_{\max}}}$	& 5 m/${s^2}$  &  ${d_0}$  & $0.6$ \\
							${{\rho_{\mathrm{sens}}}}$ 	& $0.001$  &  $\rho $ & $1.225$ kg/${{\mathrm m}^3}$ \\
							${{\rho_{\mathrm{com}}}}$ 	& $0.001$ &  $s$  & $0.05$ ${{\mathrm m}^3}$ \\	
							${R _{\min }^{{\mathrm{sec}}}}$ 	& 50 bps/Hz  &  $\Delta$ & $6$ m \\
							$\varepsilon$ 	& $0.05$  &  ${V_{\max }}$ & $20$ m/s \\
							$R_{\min }^{{\mathrm{sens}}}$ 	& $0.025$  &  $N $ & $64$ 			\\			
							\bottomrule
						\end{tabular}
					}
				\end{minipage}
			\end{table}
			
			The numerical analysis is provided to quantify the performance of the presented algorithm.
			The detailed parameter configurations are summarized in TABLE \ref{table3} \cite{ChengG2025TCOM, DanQ2025TCCN}, where $P_{U, {\mathrm{ave}}}^{{\mathrm{fly}}} = P_{J, {\mathrm{ave}}}^{{\mathrm{fly}}} = P_{{\mathrm{ave}}}^{{\mathrm{fly}}}$ and ${P_{U, \max }} = {P_{J, \max }} = {P_{\max }}$.
			To demonstrate the advantages of the proposed method (denoted as `Pro'), we compare it with the following two benchmark algorithms.
			\begin{enumerate}	
				\item  Benchmark1 (denoted as `Ben1'): Similar to \cite{LiuY2024TVT}, the trajectory of $J$, transmit power, user scheduling, and transmit beamforming design are optimized, while $U$ works with a fixed trajectory.	
				\item  Benchmark2 (denoted as `Ben2'): Similar to \cite{CaiY2018}, the trajectories of $U$ and $J$, transmit power, and user scheduling are jointly optimized, where $J$ is equipped with a single antenna.	
			\end{enumerate}
			
			The robustness of the proposed scheme is tested under the following scenarios
			\begin{itemize}
				
				\item Scenario 1: $U$ and $J$ fly from the initial position ${\bf{q}}_i^{\mathrm I} = [0, 0]^H$ to the final position ${\bf{q}}_i^{\mathrm F} = [400, 400]^H$. The square region where $E$ is distributed is centered at $[250, 125]^H$, and three users are positioned at $[100, 50]^H$, $[160, 250]^H$, and $[300, 270]^H$, respectively.	
				\item Scenario 2: $U$ and $J$ fly from the initial position ${\bf{q}}_i^{\mathrm I} = {\bf{q}}_i^{\mathrm F} = [200, 460]^H$. The square region where $E$ is distributed is centered at $[150, 250]^H$, and three users are positioned at $[350, 450]^H$, $[50, 400]^H$and $[230, 50]^H$, respectively.	
			\end{itemize}

			Fig. \ref{fig03} plots the optimized trajectories of $U$ and $J$, user scheduling, flight velocities, and the time allocation in the first scenario wherein ${D_2}$ and ${D_3}$ are located on one side of the initial trajectory, while ${D_1}$ and $E$ are on the other side. As observed from Fig. \ref{fig03a}, in the proposed scheme, $U$ moves as close as possible to the scheduled user, while $J$ proactively approaches $E$ to perform jamming. 
			$J$ in all the schemes hovers around $E$ for a certain period after approaching it. Compared with the other schemes, $J$ in Ben2 is closer to $E$ because a single antenna is used.
			As shown in Fig. \ref{fig03b}, all users are scheduled sequentially for communication, thereby ensuring fairness. 
			The results shown in Fig. \ref{fig03c} indicate that, due to the limited propulsion energy, $U$ first rapidly approaches the scheduled user and then flies at a low speed near the user to improve communication quality, while $J$ quickly approaches $E$ and then flies at a low speed near the eavesdropper to enhance the SR.
			Fig. \ref{fig03d} plots the time allocation coefficient $\eta$ in each time slot. As can be observed from Fig. \ref{fig03d}, the UAV is located far from $E$ at the initial and terminal flight periods; more time is assigned to execute the sensing task, so the values of $\eta$ corresponding to the bars on both sides are noticeably lower than those in the middle area. 
			Fig. \ref{fig03e} presents the achievable SR per time slot. It can be observed that a higher SR is achieved when $U$ is closer to the scheduled user.
			Fig. \ref{fig03f} shows that all schemes converge. It can be observed that all the schemes converge rapidly. Moreover, the ASR of the proposed scheme is higher than that of the benchmarks.
			
			Fig. \ref{fig04} presents the results for Scenario 2, in which all users are uniformly distributed around $E$. Similarly, as shown in Fig. \ref{fig04a}, $U$ in the proposed scheme flies counter-clockwise to get as close as possible to the scheduled user, and $J$ proactively approaches $E$ to perform active jamming. 
			In Ben2, since a single antenna is used, $J$ needs to be positioned extremely close to $E$, as shown in Fig. \ref{fig04c}. Meanwhile, $U$ should be placed as far away from $E$ as possible. In Ben1, as $U$ follows a fixed trajectory away from $E$ (the initial trajectory for all schemes), $J$ does not need to be in proximity to $E$. In the proposed scheme, thanks to beamforming and the joint optimization of UAV trajectories, $U$ does not need to evade $E$ with $J$'s assistance, enabling $U$ to reach the scheduled users more quickly.
			All users are scheduled in turn, as shown in Fig. \ref{fig04b}. 
			Fig. \ref{fig04c} plots the flight velocities of $U$ and $J$. Because users are distributed around $E$, the pattern in the UAV velocities is more obvious. When approaching the scheduled users, $U$ flies with a lower velocity to reduce propulsion power consumption. 
			The results in Fig. \ref{fig04d} reveal that when $U$ is closer to the user (slots 13, 41, 69), the communication time is shorter. When $E$ is closer to $U$, the time allocated for sensing is smaller.
			As shown in Fig. \ref{fig04e}, in the slots when $U$ is near the scheduled users, the curve shows pronounced, sharp peaks. Compared with Ben1, where $U$'s trajectory is fixed, the proposed scheme and Ben2 achieve an improvement in ASR.
			Fig. \ref{fig04f} shows that all schemes converge rapidly. Moreover, compared with using beamforming on $J$, optimizing the trajectory of $U$ can better enhance the secrecy performance.

			\begin{figure*}[t]
				\centering
				\subfigure[]{
					\label{fig05a}
					\includegraphics[width = \figwidThree, height = \figheiThree]{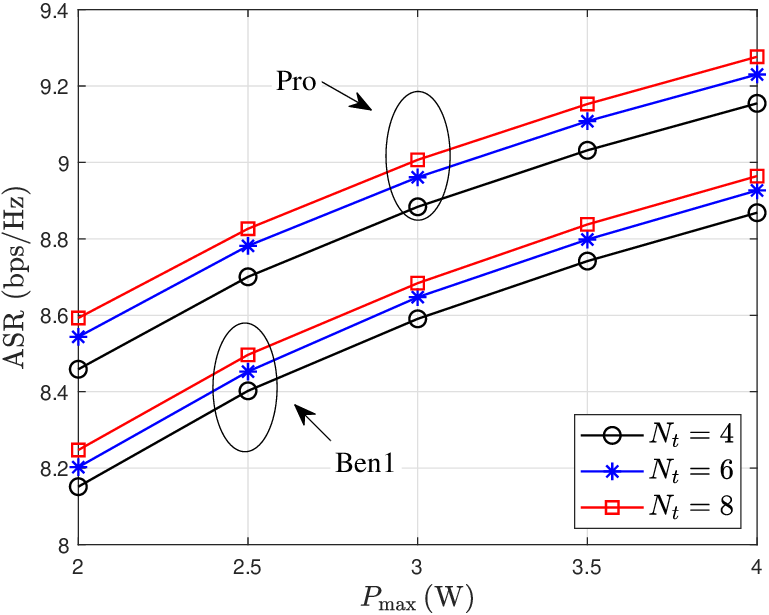}}
				\subfigure[]{
					\label{fig05b}
					\includegraphics[width = \figwidThree, height = \figheiThree]{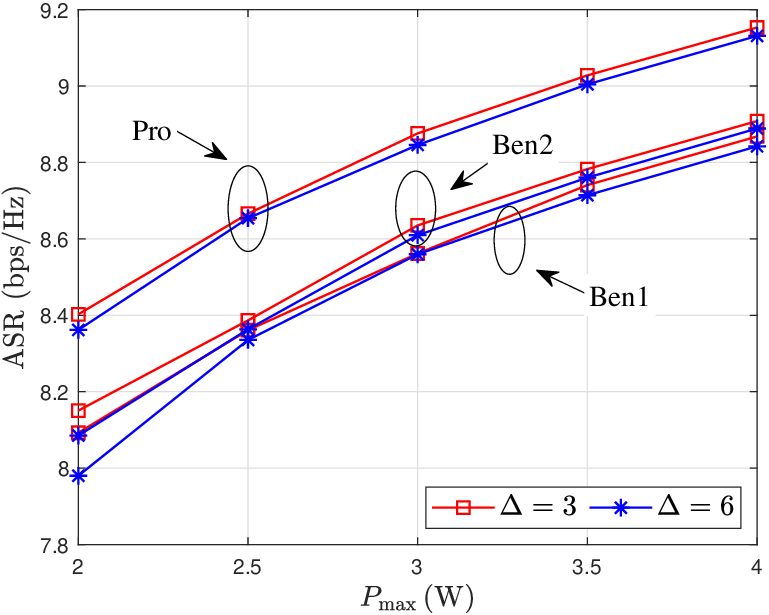}}
				\subfigure[]{
					\label{fig05c}
					\includegraphics[width = \figwidThree, height = \figheiThree]{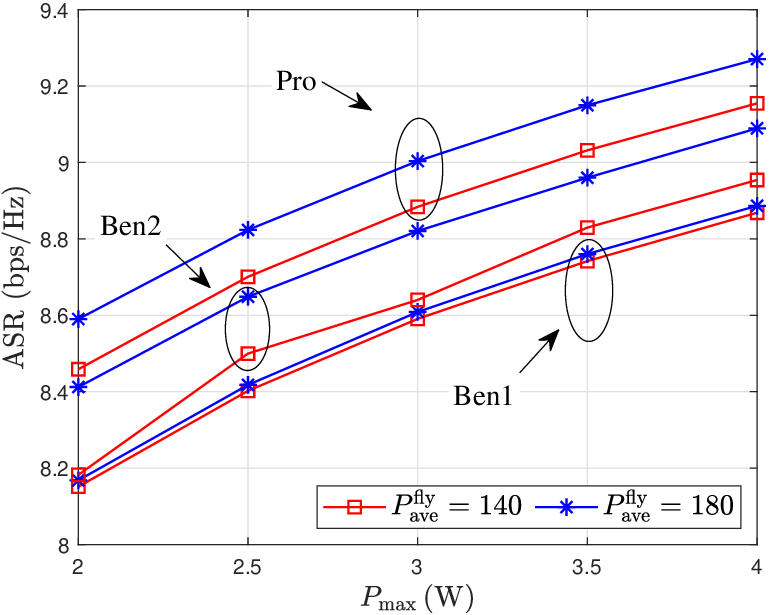}}
				\subfigure[]{
					\label{fig05d}
					\includegraphics[width = \figwidThree, height = \figheiThree]{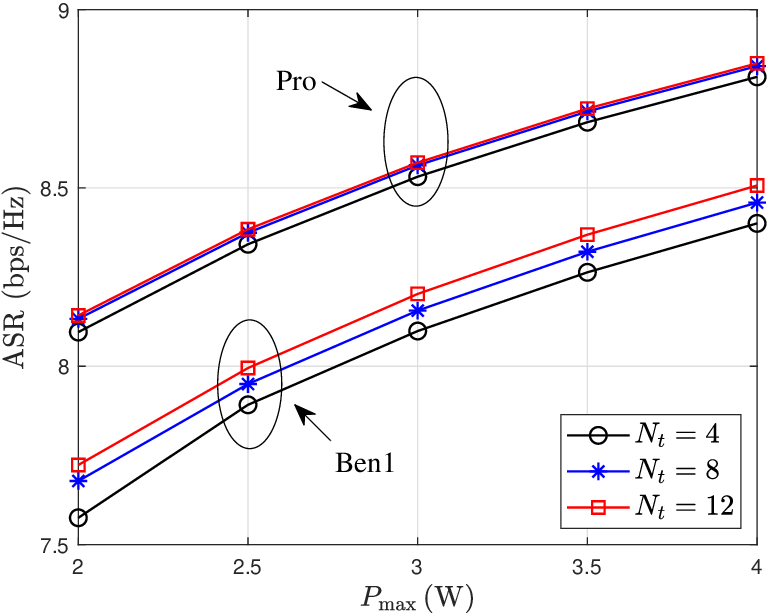}}
				\subfigure[]{
					\label{fig05e}
					\includegraphics[width = \figwidThree, height = \figheiThree]{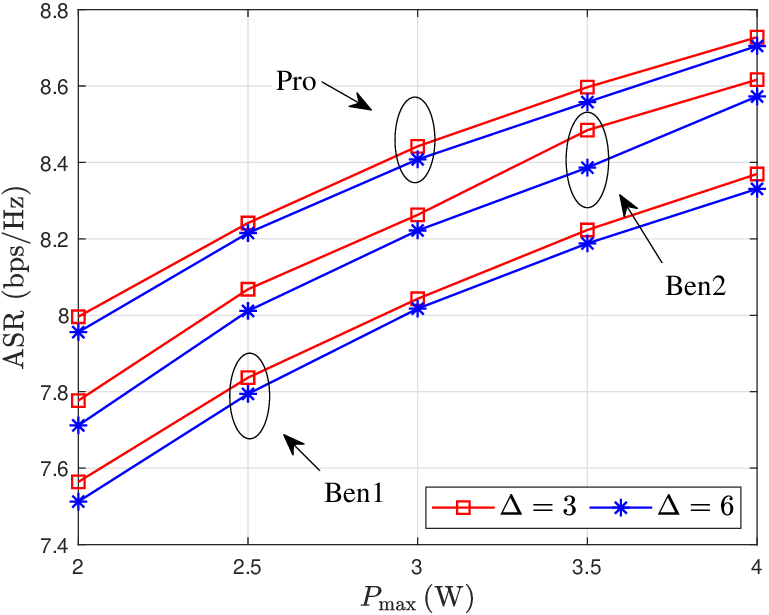}}
				\subfigure[]{
					\label{fig05f}
					\includegraphics[width = \figwidThree, height = \figheiThree]{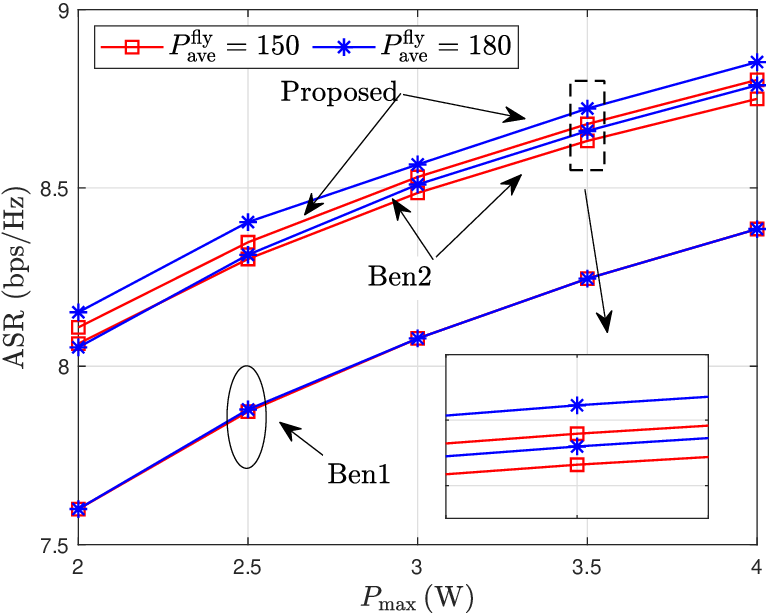}}
				\caption{ASR versus $P_{\max}$ with varying the number of transmit antennas ($N_t$), half-width of the uncertainty region ($\Delta$), and  propulsion power ($P_{\mathrm{ave}}^{{\mathrm{fly}}}$). (a)-(c) Scenario 1. (d)-(f) Scenario 2.}
				\label{fig05}
			\end{figure*}
			
			Fig. \ref{fig05} plots the ASR versus $P_{\max}$ under different numbers of transmit antennas ($N_t$), half-width of the uncertainty region ($\Delta$), and  propulsion power ($P_{\mathrm{ave}}^{{\mathrm{fly}}}$) for both scenarios. 
			As shown in Figs. \ref{fig05a} and \ref{fig05d}, ASR increases with the growth of $N_t$, since the array gain provided by the multi-antenna structure degrades the transmission condition of the eavesdropping link and weakens the signal reception capability of $E$, thereby improving the anti-eavesdropping capability of the system. 
			Figs. \ref{fig05b} and \ref{fig05e} illustrate the ASR versus $P_{\max}$ with varying $\Delta$. It can be observed that the ASR at $\Delta = 3$ m is slightly higher than at $\Delta = 6$ m, due to the more accurate location information for $E$ in the former case. As illustrated in Figs. \ref{fig05c} and \ref{fig05f}, in Scenario 1, the ASR of the proposed scheme and Ben2 increase as $P_{\mathrm{ave}}^{{\mathrm{fly}}}$ increases. However, in Scenario 2, the increase is not very significant due to the distribution of users' positions.
			Notably, the ASR of Ben1 in both scenarios is the least sensitive to variations in propulsion power. The underlying reason lies in the fixed flight trajectory adopted by $U$ in Ben1, which weakens the correlation between system performance and propulsion power.
			
			\begin{figure}[t]
				\centering
				\subfigure[]{
					\label{fig06a}
					\includegraphics[width = \figwidThree, height = \figheiThree]{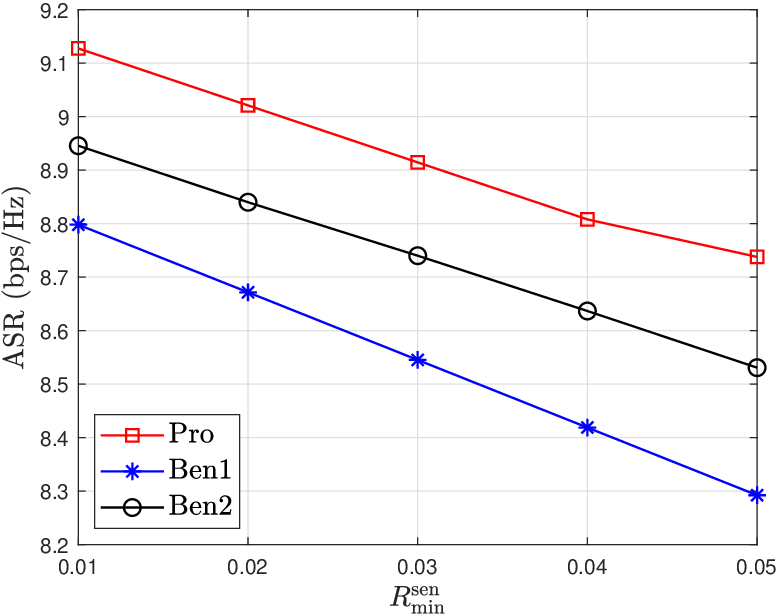}}
				\subfigure[]{
					\label{fig06b}
					\includegraphics[width = \figwidThree, height = \figheiThree]{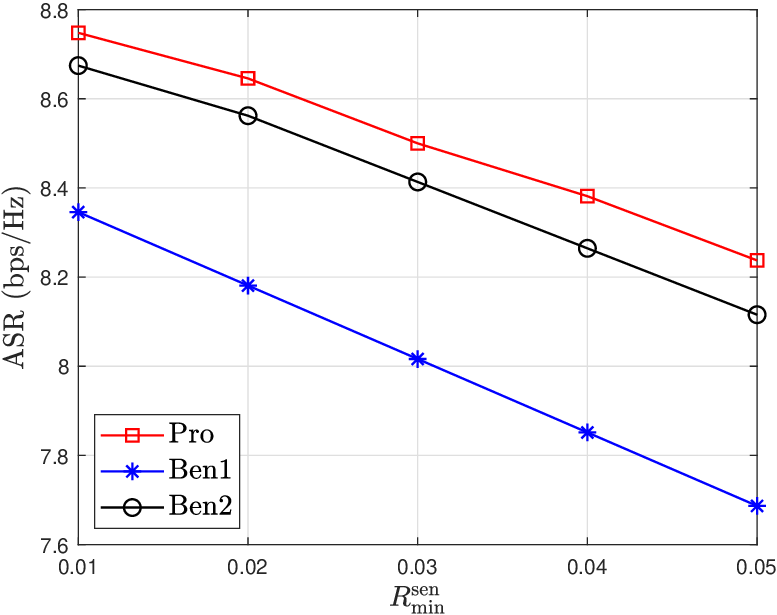}}
				\caption{ASR versus ${R_{\min }^{{\mathrm{sen}}}}$. (a) Scenario 1. (b) Scenario 2.}
				\label{fig06}
			\end{figure}
			
			As shown in Fig. \ref{fig06}, the ASR gradually decreases as $R_{\min }^{{\mathrm{sens}}}$ becomes more stringent. This is because a higher $R_{\min }^{{\mathrm{sens}}}$ compels the system to allocate more resources to meet the sensing constraint, thereby reducing the communication performance. \textit{This demonstrates the trade-off that exists between communication and sensing.} Further, it is observed that throughout the comparisons, the proposed scheme consistently outperforms the benchmark schemes in terms of ASR.

			\section{Conclusion}
			\label{sec:Conclusion}
			
			This work investigated the secrecy performance of a dual-UAV-assisted ISAC system. Within the established system framework, the aerial BS estimated the locations of the ground eavesdropper, shared the acquired sensing data with the jamming UAV, and provided downlink communication services to legitimate ground users. Meanwhile, the jamming UAV emitted AN to degrade the reception quality of potential eavesdroppers. To maximize the system ASR, we jointly optimized user scheduling, time allocation, transmit beamforming, and UAV trajectories. An efficient iterative algorithm was proposed to tackle the formulated non-convex optimization problem. Numerical results verified the convergence and practical effectiveness of the proposed algorithm. This work adopted a LoS A2G model with a single eavesdropper. A promising future research direction is to explore more complex scenarios involving multiple mobile eavesdroppers, focusing on multi-UAV cooperative sensing schemes under a probabilistic LoS channel model with 3D trajectory optimization, in which the EKF is introduced for location estimation and tracking \cite{JinH2026TCOM}. Moreover, motivated by Refs. \cite{ZhaoH2026TWC, WangD2025IoT, WangD2025IoT2, YaoJ2026TVT}, we plan to incorporate the IRS into the UAV-ISAC system. By leveraging precise phase-shift control, IRS is expected to enhance legitimate signal quality while mitigating eavesdropping risks, thereby offering dual benefits for communication and sensing.
			
			\textbf{Acknowledgments}
			
			This work was supported by the National Key Research and Development Program of China under Grant 2024YFC3306801, National Natural Science Foundation of China under Grant 62571045 and 61971080, and Natural Science Foundation of Chongqing under Grant cstc2024ycjh-bgzxm003, CSTB2025QYYJX0002, and CSTB2025NSCQ-LZX0053.
			
			\color{black}


\begin{thebibliography}{1}
				
				\bibitem{ZhangD2026Surveys}
				D. Zhang, Y. Cui, X. Cao, N. Su, Y. Gong, F. Liu, W. Yuan, X. Jing, J. Andrew Zhang, J. Xu, C. Masouros, D. Niyato, and M. Di Renzo, Integrated sensing and communications over the years: An evolution perspective, IEEE Commun. Surveys Tuts., 28 (2026) 5014-5048. 
				
				\bibitem{JinH2026npj}
				H. Jin, W. Yuan, J. Wu, J. Wang, D. Niyato, X. Wang, G. K. Karagiannidis, Z. Lin, Y. Gong, D. I. Kim, A. Petropulu, M. S. Greco, A. Jamalipour, and S. Sun, Advancing the control of low-altitude wireless networks: Architecture, design principles, and future directions, npj Wireless Technology, 2 (1) (2026) 1-8. 
				
				\bibitem{FeiZ2023Magzine}
				Z. Fei, X. Wang, N. Wu, J. Huang, and J. A. Zhang, Air-ground integrated sensing and communications: Opportunities and challenges, IEEE Commun. Mag., 61 (5) (2023) 55-61.
				
				\bibitem{MengK2024WC}
				K. Meng, Q. Wu, J. Xu, W. Chen, Z. Feng, R. Schober, and A. L. Swindlehurst, UAV-enabled integrated sensing and communication: Opportunities and challenges, IEEE Wireless Commun., 31 (2) (2024) 97-104. 
				
				\bibitem{LiuZ2025TVT}
				Z. Liu, X. Liu, J. Feng, and B. Chen, Radar estimation rate maximization for UAV assisted integrated sensing and communication, IEEE Trans. Veh. Technol., 74 (12) (2025) 19760-19765.
				
				\bibitem{JiangY2025TWC}
				Y. Jiang, X. Li, G. Zhu, K. Han, K. Meng, W. Yang, C. Liu, Q. Shi, and R. Zhang, Network-level performance analysis for air-ground integrated sensing and communication, IEEE Trans. Wireless Commun., 24 (8) (2025) 6931-6946. 
				
				\bibitem{GangY2025DCN}
				Y. Gang, Y. Zhang, and X. Wang, UAV-assisted full-duplex ISAC: Joint communication scheduling, beamforming, and trajectory optimization, Digit. Commun. Netw., 11 (5) (2025) 1628-1638. 
				
				\bibitem{KhaliliA2024TWC}
				A. Khalili, A. Rezaei, D. Xu, F. Dressler, and R. Schober, Efficient UAV hovering, resource allocation, and trajectory design for ISAC with limited backhaul capacity, IEEE Trans. Wireless Commun., 23 (11) (2024) 17635-17650. 
				
				\bibitem{ZhouL2026TWC}
				L. Zhou, C. Yang, Y. Cui, R. Zhang, Z. Wei, and Q. Shi, Joint deployment and resource allocation design for JRC-enabled multi-UAV cooperative systems, IEEE Trans. Wireless Commun., 25 (2026) 8051-8065. 
				
				\bibitem{ChengG2025TCOM}
				G. Cheng, X. Song, Z. Lyu, and J. Xu, Networked ISAC for low-altitude economy: Coordinated transmit beamforming and UAV trajectory design, IEEE Trans. Commun., 73 (8) (2025) 5832-5847. 
				
				\bibitem{YangH2025survey}
				H. Yang, Y. Liu, X. Li, Z. Bai, L. Yang, G. Pan, and H. Liu, Physical layer security and covert communication in UAV-ISAC networks: A comprehensive survey, J. King Saud Univ., Comput. Inf. Sci., 37 (10) (2025) 1-25. 
				
				\bibitem{XuY2024TWC}
				X. Yu, J. Xu, N. Zhao, X. Wang, and D. Niyato, Security enhancement of ISAC via IRS-UAV, IEEE Trans. Wireless Commun., 23 (10) (2024) 15601-15612.
				
				\bibitem{ZhangJ2024TWC}
				J. Zhang, J. Xu, W. Lu, N. Zhao, X. Wang, and D. Niyato, Secure transmission for IRS-aided UAV-ISAC networks, IEEE Trans. Wireless Commun., 23 (9) (2024) 12256-12269. 
				
				\bibitem{WangC2026TCE}
				C. Wang, X. Zhang, W. Liu, J. Ren, H. Xing, S. Wang, Y. Shen, and K. Ye, Joint beamforming and resource coordination for IRS-aided ISAC over secure LAWNs, IEEE. T. Consum. Electr, 72 (2) (2026) 3365-3377. 
				
				\bibitem{WangY2026IoT}
				Y. Wang, H. Lei, K.-H. Park, Q. Dan, X. Miao, M. A. Aboulhassan, and G. Pan, Joint trajectory and beamforming design for secure aerial ISAC systems, IEEE Internet Things J., 13 (11) (2026) 24492-24503. 
				
				\bibitem{YaoJ2025WCL}
				J. Yao, Z. Yang, Z. Yang, J. Xu, and T. Q. S. Quek, UAV-enabled secure ISAC against dual eavesdropping threats: Joint beamforming and trajectory design, IEEE Wireless Commun. Lett., 14 (10) (2025) 3199-3203.
				
				\bibitem{JinH2026IoT}
				H. Lei, H. Jin, K.-H. Park, M. A. Aboulhassan, X. Miao, and G. Pan, On secure UAV-aided ISAC system via sensing, IEEE Internet Things J., 13 (8) (2026) 16680-16692. 
				
				\bibitem{JinH2026TCOM}
				H. Lei, H. Jin, K.-H. Park, J. Ye, L. Yang, G. Pan, and Y. Li, On secure EKF-enhanced UAV-ISAC systems, IEEE Trans. Commun., 74 (2026) 11964-11976. 
				
				\bibitem{JiangC2025IoT}
				H. Lei, C. Jiang, K.-H. Park, M. A. Aboulhassan, S. Zhou, and G. Pan, On secure UAV-aided ISCC systems, IEEE Internet Things J., 12 (19) (2025) 40851-40862. 
				
				\bibitem{LouY2026TVT}
				L. Lou, Y. Liu, F. Foukalas, H. Lei, G. Pan, T. A. Tsiftsis, and H. Liu, Maneuverable-jamming-aided secure communication and sensing in A2G-ISAC systems, IEEE Trans. Veh. Technol., 75 (7) (2026) 13800-13814. 
				
				\bibitem{LiuY2024TVT}
				Y. Liu, X. Liu, Z. Liu, Y. Yu, M. Jia, Z. Na, and T. S. Durrani, Secure rate maximization for ISAC-UAV assisted communication amidst multiple eavesdroppers, IEEE Trans. Veh. Technol., 73 (10) (2024) 15843-15847. 
				
				
				\bibitem{LeiH2024TCCN}
				H. Lei, X. Wu, K.-H. Park, and G. Pan, 3D trajectory design for energy-constrained aerial CRNs under probabilistic LoS channel, IEEE Trans. Cogn. Commun. Netw., 11 (3) (2025) 1522-1534. 
				
				\bibitem{LeiH2026TCCN}
				H. Lei, X. Wu, K.-H. Park, G. Chen, and G. Pan, Joint 3-D trajectory design and resource allocation for secure dual-UAV-aided underlay systems, IEEE Trans. Cogn. Commun. Netw., 12 (2026) 9147-9161. 
				
				\bibitem{MHua2024TWC}
				M. Hua, Q. Wu, W. Chen, O. A. Dobre, and A. L. Swindlehurst, Secure intelligent reflecting surface-aided integrated sensing and communication, IEEE Trans. Wireless Commun., 23 (1) (2024) 575-591.
				
				\bibitem{LuW2022TCOM}
				W. Lu, Y. Ding, Y. Gao, Y. Chen, N. Zhao, Z. Ding, and A. Nallanathan, Secure NOMA-based UAV-MEC network towards a flying eavesdropper, IEEE Trans. Commun., 70 (5) (2022) 3364-3376. 
				
				\bibitem{DingY2023JSTSP}
				Y. Ding, Y. Feng, W. Lu, S. Zheng, N. Zhao, L. Meng, A. Nallanathan, and X. Yang, Online edge learning offloading and resource management for UAV-assisted MEC secure communications, IEEE J. Sel. Topics Signal Process., 17 (1) (2023) 54-65. 
				
				\bibitem{DanQ2025TVT}
				Q. Dan, H. Lei, K.-H. Park, G. Pan, and M.-S. Alouini, Two birds with one stone: Beamforming design for target sensing and proactive eavesdropping, IEEE Trans. Veh. Technol., 75 (8) (2026) 17590-17603.
				
				\bibitem{DingX2026TVT}
				X. Ding, Q. Lu, Y. Zhang, G. Li, X. Gao, N. Ye, D. Niyato, and K. Yang, Few-shot recognition and classification framework for jamming signal: A CGAN-based fusion CNN approach, IEEE Trans. Veh. Technol., 75 (8) (2026) 17494-17508. 
				
				\bibitem{NingZ2026JSAC}
				Z. Ning, L. Li, X. Wang, H. Lei, Y. Huang, L. Guo, and Y. Zhang, Joint beamforming and trajectory design for multi-UAV assisted integrated sensing, communication, and power transfer networks, IEEE J. Sel. Areas Commun., 44 (2026) 5282-5298. 
				
				\color{black}
				\bibitem{LiuZ2024TWC}
				Z. Liu, X. Liu, Y. Liu, V. C. M. Leung, and T. S. Durrani, UAV assisted integrated sensing and communications for internet of things: 3D trajectory optimization and resource allocation, IEEE Trans. Wireless Commun., 23 (8) (2024) 8654-8667. 
				
				\bibitem{JiangL2026IoT}
				H. Lei, L. Jiang, K.-H. Park, Q. Dan, X. Miao, and Y. Li, Covert communication in dual-UAV-aided ISAC systems with collusive wardens, IEEE Internet Things J. 13 (18) (2026) 43225-43237. 
				
				\bibitem{DanQ2025TCCN}
				Q. Dan, H. Lei, K.-H. Park, and G. Pan, Beamforming for secure RSMA-aided ISAC systems, IEEE Trans. Cogn. Commun. Netw., 11 (5) (2025) 2970-2983. 
				
				\bibitem{LvL2019}
				L. Lv, F. Zhou, J. Chen, and N. Al-Dhahir, Secure cooperative communications with an untrusted relay: A NOMA-inspired jamming and relaying approach, IEEE Trans. Inf. Forensics Security, 14 (12) (2019) 3191-3205. 
				
				\bibitem{CaiY2018}
				Y. Cai, F. Cui, Q. Shi, M. Zhao, and G. Y. Li, Dual-UAV-enabled secure communications: Joint trajectory design and user scheduling, IEEE J. Sel. Areas Commun., 36 (9) (2018) 1972-1985. 
				
				\bibitem{XingH2016}
				H. Xing, L. Liu, and R. Zhang, Secrecy wireless information and power transfer in fading wiretap channel, IEEE Trans. Veh. Technol., 65 (1) (2016) 180-190. 
				
				\bibitem{ZhongC2019CL}
				C. Zhong, J. Yao, and J. Xu, Secure UAV communication with cooperative jamming and trajectory control, IEEE Commun. Lett., 23 (2) (2019) 286-289. 
				
				
				\bibitem{LiuX2024IOT}
				X. Liu, Y. Liu, Z. Liu, and T. S. Durrani, Fair integrated sensing and communication for multi-UAV enabled internet of things: Joint 3D trajectory and resource optimization, IEEE Internet Things J., 11 (18) (2024) 29546-29556.
				
				\bibitem{ZengY2019TWC}
				Y. Zeng, J. Xu, and R. Zhang, Energy minimization for wireless communication with rotary-wing UAV, IEEE Trans. Wireless Commun., 18 (4) (2019) 2329-2345. 
				
				\bibitem{LeiH2023IoT}
				H. Lei, H. Yang, K.-H. Park, I. S. Ansari, J. Jiang, and M.-S. Alouini, Joint trajectory design and user scheduling for secure aerial underlay IoT systems, IEEE Internet Things J., 10 (15) (2023) 13637-13648. 
				
				
				\bibitem{LyuZ2023TWC}
				Z. Lyu, G. Zhu, and J. Xu, Joint maneuver and beamforming design for UAV-enabled integrated sensing and communication, IEEE Trans. Wireless Commun., 22 (4) (2023) 2424-2440. 
				
				\bibitem{ZhaoH2026TWC}
				H. Zhao, M. Wu, D. Wang, M. Guizani, and V. C. M. Leung, Sensing-assisted secure beamforming for IRS-enabled ISAC with leakage suppression, IEEE Trans. Wireless Commun., 25 (2026) 17755-17769. 
				
				\bibitem{WangD2025IoT}
				D. Wang, T. Liu, L. Li, W. Yang, Y. Jin, H. Zhao, Y. He, and R. Zhang, Performance analysis of UAV-IRS assisted short-packet secure communications, IEEE Internet Things J., 12 (24) (2025) 53039-53054.
				
				\bibitem{WangD2025IoT2}
				D. Wang, J. Li, Q. Lv, Y. He, L. Li, Q. Hua, O. Alfarraj, and J. Zhang, Integrating reconfigurable intelligent surface and AAV for enhanced secure transmissions in IoT-enabled RSMA networks, IEEE Internet Things J., 12 (8) (2025) 9405-9419. 
				
				\bibitem{YaoJ2026TVT}
				J. Yao, Z. Dai, J. Xu, Y. Fang, G. Han, and T. Q. S. Quek, Secrecy rate maximization for coupled phase-shift STAR-IRS-assisted ISAC networks under dual eavesdropping attacks, IEEE Trans. Veh. Technol., 75 (2) (2026) 3452-3457.
	
\end{thebibliography}
\end{document}